\documentclass{article}

\usepackage{authblk}
\usepackage{lmodern}
\usepackage[T1]{fontenc}
\usepackage[utf8]{inputenc}
\usepackage[margin=0.9in]{geometry}
\usepackage{amssymb,amsmath,amsthm,amsfonts}

\usepackage{xcolor}
\usepackage{soul} 
\usepackage{tikz}
\usepackage{pgfplots}
\usepackage{makecell}
\usepackage[inline]{enumitem}
\usepackage{subcaption}
\usepackage{graphicx}
\usepackage{wrapfig}
\usepackage{amsmath}
\usepackage{amssymb}
\usepackage{mathtools}
\usepackage{multirow}
\usepackage{balance}
\usepackage{tablefootnote}

\usepackage{pifont}

\usetikzlibrary{arrows,shapes,positioning,shadows,trees}

\tikzset{
  basic/.style  = {draw, text width=2cm, drop shadow, font=\sffamily, rectangle},
  root/.style   = {basic, rounded corners=2pt, thin, align=center,
                   fill=green!30},
  level 2/.style = {basic, rounded corners=6pt, thin,align=center, fill=green!60,
                   text width=8em},
  level 3/.style = {basic, thin, align=left, fill=pink!60, text width=6.5em}
}

\usepackage[pagebackref]{hyperref}
\hypersetup{
    colorlinks,
    linkcolor={red!50!black},
    citecolor={blue!50!black},
    urlcolor={blue!80!black},
    pdfstartview={XYZ null null 1.25}
}
\usepackage{orcidlink}
\providecommand{\keywords}[1]
{
  \textbf{\textit{Keywords---}} #1
}

\def\mySectionSymbol~{\S{}}

\usepackage{abstract}

\begin{document}
\title{Keystroke Dynamics: Concepts, Techniques, and Applications}

\newcommand{\clrksn}{{*}}
\newcommand{\rit}{{$^\mathsection$}}
\newcommand{\afrl}{{$^{\ddagger}$}}

\renewcommand*{\Authsep}{\ }
\renewcommand*{\Authand}{\ }
\renewcommand*{\Authands}{\ }

\author[\clrksn]{Rashik Shadman\orcidlink{0009-0007-6470-9042}}
\author[\clrksn]{Ahmed Anu Wahab\orcidlink{0000-0003-4677-5269}}
\author[\afrl]{Michael Manno\orcidlink{0009-0003-0536-4276}}
\author[\clrksn]{Matthew Lukaszewski\orcidlink{0009-0000-3485-8062}} 
\author[\rit]{\authorcr Daqing Hou\orcidlink{0000-0001-8401-7157}} 
\author[\clrksn]{Faraz Hussain\orcidlink{0000-0001-8971-1850}} 
\affil[\clrksn]{Clarkson University, Potsdam, NY, USA\authorcr {\tt \{shadmar,wahabaa,lukaszmp,fhussain\}@clarkson.edu}\vspace{0.4em}}
\affil[\afrl]{Air Force Research Laboratory, Rome, NY, USA\authorcr {\tt michael.manno@us.af.mil}\vspace{1em}}
\affil[\rit]{Rochester Institute of Technology, Rochester, NY, USA\authorcr{\tt dqvse@rit.edu}\vspace{0.4em}}

\date{}
\maketitle

\newcommand{\subsubsubsection}[1]{\paragraph{#1}\mbox{}\\}
\setcounter{secnumdepth}{4}
\setcounter{tocdepth}{4}
\addtolength{\itemsep}{-0.1in}
\addtolength{\topsep}{-0.07in}
\addtolength{\textfloatsep}{-0.05in}
\addtolength{\partopsep}{-0.03in}
\addtolength{\parskip}{-0.02in}

\begin{abstract}
 \normalsize
 
Reliably identifying and verifying subjects remains integral to computer system security. Various novel authentication techniques, such as biometric authentication systems, have been developed in recent years. This paper provides a detailed review of keystroke-based authentication systems and their applications.  Keystroke dynamics is a behavioral biometric that is emerging as an important tool for cybersecurity as it promises to be nonintrusive and cost-effective. In addition, no additional hardware is required, making it convenient to deploy.  This survey covers novel keystroke datasets, state-of-the-art keystroke authentication algorithms,  keystroke authentication on touch screen and mobile devices, and various prominent applications of such techniques beyond authentication. The paper covers all the significant aspects of keystroke dynamics and can be considered a reference for future researchers in this domain. 

The paper includes a discussion of the latest keystroke datasets, providing researchers with an up-to-date resource for analysis and experimentation. In addition, this survey covers the state-of-the-art algorithms adopted within this domain, offering insights into the cutting-edge techniques utilized for keystroke analysis. Moreover, this paper explains the diverse applications of keystroke dynamics, particularly focusing on security, verification, and identification uses. Beyond these crucial areas, we mention additional applications where keystroke dynamics can be applied, broadening the scope of understanding regarding its potential impact across various domains. Unlike previous survey papers, which typically concentrate on specific aspects of keystroke dynamics, our comprehensive analysis presents all relevant areas within this field. By introducing discussions on the latest advances, we provide readers with a thorough understanding of the current landscape and emerging trends in keystroke dynamics research. Furthermore, this paper presents a summary of future research opportunities, highlighting potential areas for exploration and development within the realm of keystroke dynamics. This forward-looking perspective aims to inspire further inquiry and innovation, guiding the trajectory of future studies in this dynamic field.
\end{abstract}


\section{INTRODUCTION}
User authentication is a process that verifies that someone who is attempting to access services and applications is who they claim to be. There are three broad categories of approaches
to authenticating a user. The first category is based on \emph{What you know}, such as the username and password or answers to security questions. This is also known as Knowledge-Based
Authentication (KBA). The second category is based on \emph{What you have}, i.e., a physical object that can be possessed, for example, a security badge or a security key such as Yubikey. The third category is based on \emph{Who you are}, which can be i) \emph{physiological}, e.g., face, fingerprint, or iris, 
ii) \emph{behavioral}, e.g., keystroke dynamics, mouse dynamics, mobile motion or swipes. 
Although KBA is currently still the dominant approach for identity management, there are active efforts in creating alternatives such as face recognition and FIDO\footnote{FIDO stands for Fast IDentity Online. It's a set of open authentication standards designed to make online logins more secure and user-friendly by reducing reliance on passwords.}. In particular,
an emerging field of user authentication is behavioral biometrics, which authenticates a user based on their behavior when interacting with a computing device.

\begin{wrapfigure}[25]{r}{0.5\textwidth}
  \centering
  \fbox{\includegraphics[width=0.5\textwidth]{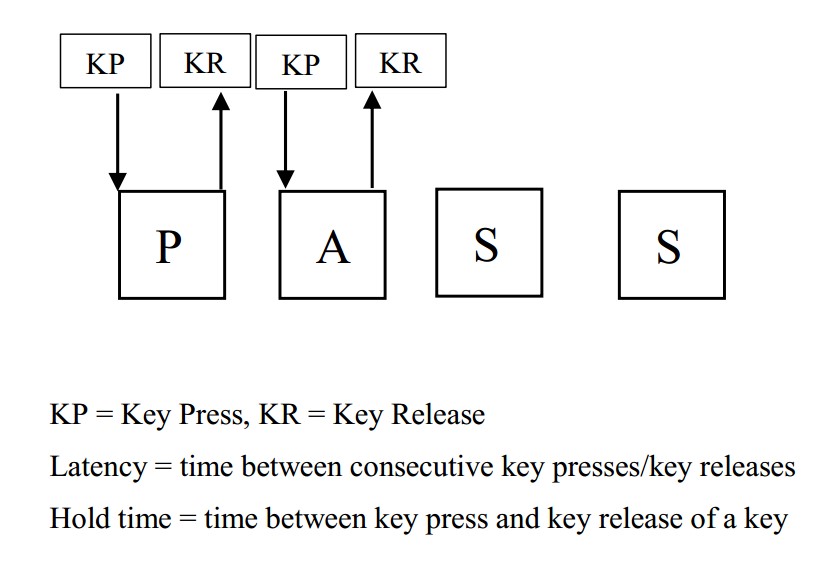}}
  \caption{Timing Features of Keystrokes. Latency and hold time: these timing features are used to create a template of a user and at the time of testing these features are compared to verify the user. Between two consecutive keystrokes, there can be 6 timing features: hold time of two keys, time between key press of first key and key press of second key = KP2 - KP1, time between key press of first key and key release of second key = KR2 - KP1, time between key release of first key and key press of second key = KP2 - KR1, time between key release of first key and key release of second key = KR2 - KR1.}
  \label{fig:Keystroke_feature}
\end{wrapfigure}

Keystroke-based biometric authentication systems are a convenient and user-friendly authentication method in which a person is authenticated based on their typing pattern. Although such systems have great potential, their application in the real world so far has been limited. Keystroke biometric authentication systems have several advantages: 
\begin{enumerate}
 \item They are cost-effective as no additional hardware is required; only the regular keyboard is needed. 
 \item They are non-intrusive and do not create any extra hassle for the user. 
 \item The systems can be deployed remotely.
\end{enumerate}
Keystroke authentication operates by creating a template for each user based on their typing pattern during an enrollment period. Once enrolled,
the test sample from the user is contrasted with the representation of the same user and a matching score will be calculated. Matching scores are calculated on the basis of the timing features of keystrokes. \autoref{fig:Keystroke_feature} provides an overview of the timing features of keystrokes.

Researchers use three types of keystroke data for authentication:
\begin{enumerate}
 \item Free-text, also known as dynamic text, allows the user to type freely without any restriction \cite{murphy2017shared}. For example, if a user writes a paragraph on a topic, it will be free-text. Free-text based keystroke dynamics can enable continuous authentication as a user types. 
 \item Fixed-text or static text \cite{killourhy2009comparing}. Fixed-text is constant during the authentication process, during both template creation and testing. The keystroke data for a password can be used to create a second authentication factor.
 \item Semi fixed-text shares some characteristics with free and fixed-text. Examples of semi fixed-text include Linux commands \cite{wang2024keystroke} and the webform keystroke data \cite{wahab2021utilizing}.
\end{enumerate}

\subsection{Justification for a New Survey}
The oldest survey article we reviewed is Alsultan and Warwick's article from 2013 \cite{alsultan2013keystroke} that focused on free-text keystroke systems. Fixed-text or semi fixed-text keystroke systems were not included in their survey. In our survey, we focus on all three types of keystroke systems. Zhong and Deng \cite{zhong2015survey} presented a comprehensive survey of keystroke dynamics; Giot, Dorizzi, and Rosenberger \cite{giot2015review} provided a review of previously created keystroke benchmarking datasets; Ali et al. \cite{ali2015authentication} presented an elaborate survey of the keystroke data collection details and classification algorithms. However, as these surveys were published in 2015, they do not cover recent trends in keystroke dynamics research or newer benchmark datasets, e.g., \cite{cceker2016user}, \cite{sun2016shared}, \cite{murphy2017shared}, \cite{wahab2021utilizing}, \cite{wahab2022shared}. 

Saini, Kaur, and Bhatia \cite{saini2016keystroke} and Saifan et al. \cite{saifan2016survey} presented a comprehensive review on keystroke dynamics research, especially for touch screen and mobile devices. However, these papers do not cover recent developments as they were published in 2016. Also, these papers focused only on mobile devices.

Shinde, Shetty, and Mehra \cite{shinde2016survey} mainly focused on different aspects of static-text keystroke authentication. They did not discuss dynamic or continuous keystroke authentication in their survey. In our survey, we explain both static and dynamic keystroke authentication. Ali et al. \cite{ali2017keystroke} provided a comprehensive assessment of the newest datasets and algorithms used in keystroke dynamics research. This survey was conducted in 2017, and new datasets and algorithms are now available. 

Sanghi and Arya \cite{sanghi2017survey} presented a review of the methods and metrics used in keystroke dynamics. This paper described various applications of keystroke dynamics as well. Sadikan, Ramli, and Fudzee \cite{sadikan2019survey} performed a comprehensive analysis of the applications of keystroke dynamics. In this paper, we describe the latest applications of keystroke dynamics. Tewari \cite{tewari2022keystroke} discussed only deep learning algorithms for keystroke dynamics. This paper provides a more comprehensive description of various types of keystroke dynamics algorithms. 

To summarize, this survey covers keystroke dynamics performance metrics, the latest benchmark datasets for keystroke dynamics, state-of-the-art keystroke authentication algorithms, keystroke feature engineering techniques, keystroke dynamics for touch screen and mobile devices, and the most important applications of keystroke dynamics.

\begin{table}
  \centering
  \scriptsize
  \caption{A summary of recent survey papers focusing on keystroke dynamics. The surveys are compared across five categories based on whether they cover various areas, viz. datasets, algorithms, mobile device keystroke capture, feature engineering approaches, and applications.}
  \label{table:ComparisonSurvey}
  \setlength{\tabcolsep}{5pt}
  \renewcommand{\arraystretch}{2}
  \centering
  \resizebox{\textwidth}{!}{%
  \begin{tabular}{|m{4.5cm}|m{1cm}|m{1.2cm}|m{1.4cm}|m{1.5cm}|m{1.3cm}|m{1.5cm}|m{1.6cm}|}  
   \hline
    {} & {} & {} & \multicolumn{5}{c|}{Covered Areas}\\\cline{4-8}
    {\textbf{Survey Paper}} & \makecell{\textbf{Year}} & {\textbf{Citations}} & {\textbf{Datasets}} & {\textbf{Algorithms}} & {\textbf{Mobile Devices}} & {\textbf{Feature Engineering}} & {\textbf{Applications}}\\
    \hline
    {Alsultan and Warwick \cite{alsultan2013keystroke}} & {2013} & {115} & {Yes} & {Yes} & {No} & {No} & {Yes}\\
    \hline
    {Zhong and Deng \cite{zhong2015survey}} & {2015} & {60} & {Yes} & {Yes} & {Yes} & {No} & {Yes}\\
    \hline
    {Giot, Dorizzi, and Rosenberger \cite{giot2015review}} & {2015} & {43} & {Yes} & {No} & {No} & {No} & {No}\\
    \hline
    {Ali et al. \cite{ali2015authentication}} & {2015} & {30} & {Yes} & {Yes} & {No} & {No} & {No}\\
    \hline
    {Saini, Kaur, and Bhatia \cite{saini2016keystroke}} & {2016} & {20} & {No} & {No} & {Yes} & {No} & {No}\\
    \hline
    {Saifan et al. \cite{saifan2016survey}} & {2016} & {11} & {No} & {No} & {Yes} & {No} & {No}\\
    \hline
    {Shinde, Shetty and Mehra \cite{shinde2016survey}} & {2016} & {7} & {Yes} & {Yes} & {No} & {No} & {No}\\
    \hline
    {Ali et al. \cite{ali2017keystroke}} & {2017} & {103} & {Yes} & {Yes} & {No} & {No} & {No}\\
    \hline
    {Sanghi and Arya \cite{sanghi2017survey}} & {2017} & {2} & {No} & {Yes} & {No} & {No} & {Yes}\\
    \hline
    {Sadikan, Ramli, and Fudzee \cite{sadikan2019survey}} & {2019} & {10} & {Yes} & {No} & {No} & {No} & {Yes}\\
    \hline
    {Maiorana, Kalita, and Campisi \cite{maiorana2021mobile}} & {2021} & {4} & {No} & {No} & {Yes} & {No} & {No}\\
    \hline      
    {Tewari \cite{tewari2022keystroke}} & {2022} & {1} & {Yes} & {Yes} & {No} & {No} & {No}\\
    \hline                       
    {This Survey} & {} & {} & {Yes} & {Yes} & {Yes} & {Yes} & {Yes}\\
    \hline
  \end{tabular}}
\end{table}

\autoref{table:ComparisonSurvey} shows a comparison of existing surveys on keystroke dynamics. We highlight their unique contributions, advantages, and disadvantages. We have categorized the studies based on their contribution to one of the following categories: keystroke datasets, keystroke authentication algorithms, 
keystroke feature engineering approaches, keystroke authentication on touch screen and mobile devices, and applications of keystroke dynamics.

\subsection{Methodology}
\begin{wrapfigure}[28]{r}{0.6\textwidth}
\center
\fbox{\includegraphics[width=0.6\textwidth]{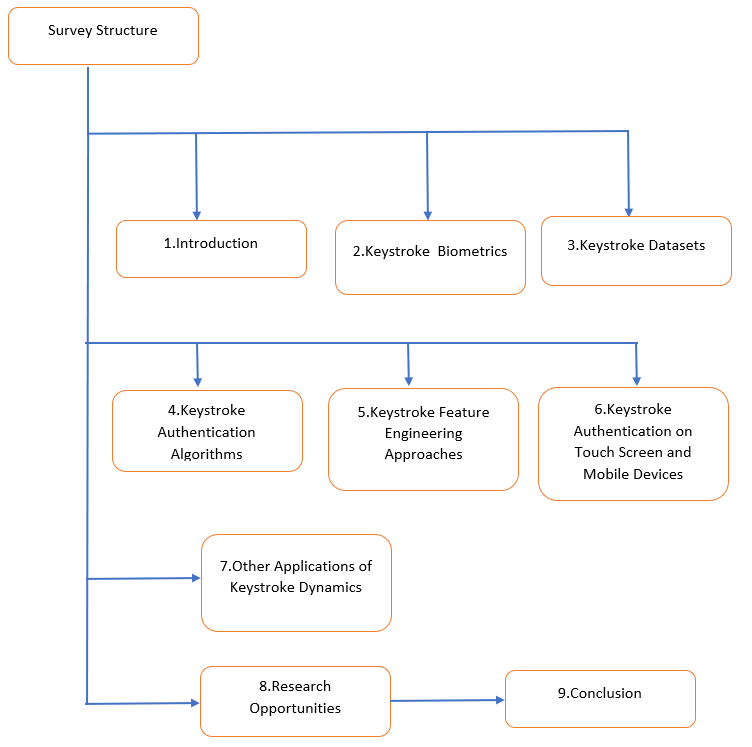}}
  \caption{An overview of the structure of this survey.}
  \label{fig:Structure}
\end{wrapfigure}

Through this paper, we comprehensively cover recent developments in keystroke dynamics research and their applications. We collected papers from 2013-present on keystroke dynamics from sources such as Google Scholar, IEEE, and Academia.edu. The following search terms were used:
\begin{itemize}
 \item keystroke dynamics
 \item keystroke dataset/database
 \item keystroke authentication algorithm
 \item keystroke dynamics + mobile devices, keystroke dynamics + touch screen devices
\end{itemize}
We categorized them based on their contribution and uniqueness. The following contributions were considered:
\begin{itemize}
 \item Novel keystroke dataset/database
 \item New methods to improve the authentication performance of keystroke dynamics-based systems such as novel/efficient keystroke authentication algorithm or software, effective keystroke feature engineering approach
 \item Novel keystroke authentication technique based on touch screen and mobile devices
 \item Potential application of keystroke dynamics
 \item Future research problem/direction in the field of keystroke dynamics
\end{itemize}
If a paper contributed in any of the above areas, we have included it in this survey. Predominantly, papers published after 2013 were considered for this survey as we wanted to cover the developments of keystroke dynamics during the last decade.
Previous surveys on keystroke dynamics focused on specific areas such as keystroke dynamics on touch screen and mobile devices, keystroke authentication algorithms, and keystroke datasets. An overview of the structure of this paper is shown in \autoref{fig:Structure}.

\vspace{-5pt}

\section{Keystroke Biometrics}
Keystroke dynamics research has been ongoing for decades and the main intended application has been verification and identification of subjects. 
Modern computer systems contain vital user information making it important to ensure the 
security of computer systems. Authentication via something a user knows (i.e. a password) is no longer sufficient to guarantee the security of sensitive data. 

In the context of keystroke dynamics, authentication, verification, and identification have different goals. Authentication means ensuring that a user is who they claim to be by 
comparing their typing patterns with a previously stored profile. In order to confirm the access of the correct individual, authentication is used when logging into a system. Verification can be done during different stages of interaction. In order to detect unauthorized/imposter access continuously during a session, verification is performed. Identification
seeks to determine the identity of an unknown user from a pool of candidates based on their typing patterns. In the identification process, the user input is effectively matched to the most similar
profile within a database. 

\subsection{Security and Authentication}
Sridhar, Vaidya and Yawalkar \cite{sridhar2015intrusion}
proposed a keystroke authentication system for intruder detection in secure virtual environments. For this purpose, 
authentication data from login attempts was stored. Feature sets were extracted from 200 samples of username and password data. Fuzzy Logic algorithms were used in order to develop signatures for each participant. 
The typing pattern of each participant was compared to the respective signature in order to authenticate the participants. Membership
functions of fuzzy logic were implemented for authenticating a participant. For 200 samples, they achieved a FAR of 0.0\% and a FRR of
0.0\% for both Bell MF and Gaussian MF. 

\subsection{Person Identification and Forensic Investigation}
Mondal and Bours \cite{mondal2017person} investigated the usefulness of keystroke dynamics in identifying a person. They proposed a pairwise user coupling (PUC) technique along with different machine learning algorithms. They also analyzed person
identification via an optimized feature set. For their experiment, they used a novel keystroke dataset collected from three online exams. 
The subjects were PACE University undergraduates. 
64 students participated in the data collection. The subjects typed a minimum of 500 keystrokes in each of the three exams. Mondal and Bours used a subset of data from the first exam as a training set. The remaining data was used for testing. They implemented the PUC technique on their dataset using machine
learning classifiers. The best accuracy achieved was 89.7\%.

Mohlala, Ikuesan, and Venter \cite{mohlala2017user} proposed a method based on keystroke dynamics to collect evidence for digital forensic readiness. The main goal was to collect 
solid keystroke evidence for forensic investigations. Their model consisted of two core phases: 1) the pattern development phase and 2) the pattern testing/validation phase.
In the development phase, input data is taken
from the known user and continually developed by applying the data preprocessing, machine learning, and pattern extraction subprocesses before storing the data in the database.
They used an Android application with a software keyboard to record raw data. The dataset consisted of 42 subjects. The subjects typed fixed password with 51 samples per subject. For the experiment process, Mohlala, Ikuesan, and Venter implemented their developed tool on the benchmark dataset \cite{antal2015evaluation}. Each subject's pattern was accurately attributed to the respective user on the test set, except for one user. However, all subjects were accurately attributed using the attribution mechanism. Based on the results, it was clear that this study could achieve reliable accuracy for
subject attribution in digital forensics.

\subsection{User Authentication for Online Assessments and Examinations}
Chen et al. \cite{chen2021keystroke} designed a keystroke dynamics-based user authentication system for online assessments and examinations. Their framework is based on
an edge computing architecture. There were keystroke profiles of the users in the framework. In this study, Chen et al. used four features: hold time (HT), press flight time (FT),
HT proportion (HTP), and FT proportion (FTP). There were two parts in the profile for static authentication and continuous authentication respectively. 
In order to perform static keystroke authentication, they used an anomaly detector based on the Gaussian model. In the case of
continuous keystroke authentication, keystroke events were collected continuously by the framework in order to form a continuous data stream of extracted keystroke timing features. 
For evaluation, three public datasets were used (\cite{killourhy2009comparing}, \cite{giot2012web}, \cite{sun2016shared}).
To investigate the effectiveness of the framework, they used a real-world case of an online examination. In order to verify their basic detector, they used the dataset from Killourhy and Maxion \cite{killourhy2009comparing} and achieved an EER of 6.62\%. To evaluate the performance of static keystroke authentication, they used the dataset from Giot et al. \cite{giot2012web} and achieved an EER of 5.71\%. EER could be decreased to 4.03\% using a
dynamic user profiles mechanism. To evaluate the performance of continuous keystroke authentication, they used the dataset from Sun et al. \cite{sun2016shared} and achieved an average EER
of 2\%. The edge computing architecture was implemented to develop a prototype of an online examination system. There were three phases in the online examination system:
registration phase, login phase, and testing phase. In order to simulate possible security-related situations, 27 users participated in the case study. The obtained results for
this case study were promising. 

\subsection{Keystroke Dynamics Performance Metrics}
There are a number of performance metrics used in keystroke dynamics, and each is used to present some specific information about the biometric system. This section describes all the metrics and when they are preferably used.

\subsubsection{False Accept Rate and False Reject Rate}
The most common performance metrics used in biometric systems are the False Accept Rate (FAR) and the False Reject Rate (FRR). The FAR is the measure of the likelihood that the biometric system will incorrectly grant access to an unauthorized user \cite{dorca2017identifying}. This is calculated as the ratio of the number of false acceptances divided by the total number of impostor attempts, as shown in equation \ref{equation:FAR}.

The measure of the likelihood that an authorized user will be incorrectly rejected by the biometric system is called the FRR. This is measured by dividing the number of false rejections by the total number of genuine attempts, as shown in Equation \ref{equation:FRR}.

The ideal value for both FAR and FRR is 0\%, which translates to granting access to all genuine users while all impostors are caught. However, real-world biometric systems are not ideal. Therefore, for both FAR and FRR, a value as close to zero as possible is desired. A system with 0\% FAR is considered to have the utmost security, while a system with 0\% FRR is highly convenient. There is usually a trade-off between security and convenience, as a highly secure system may reject too many genuine users, thus being inconvenient, while a highly convenient system may be insecure and therefore ineffective.
The FAR is also known as the False Positive Rate (FPR), while the FRR = 1 - True Positive Rate (TPR).

\begin{equation}
\label{equation:FAR}
FAR = \frac{\text{Number of false acceptances}}{\text{Total number of impostor attempts}}
\end{equation}

\begin{equation}
\label{equation:FRR}
FRR = \frac{\text{Number of false rejections}}{\text{Total number of genuine attempts}}
\end{equation}

\subsubsection{Equal Error Rate}
\begin{figure}[t]
\centering
   \begin{subfigure}[b]{0.46\textwidth}
   \centering
   \fbox{\includegraphics[width=\textwidth, height=6.5cm]{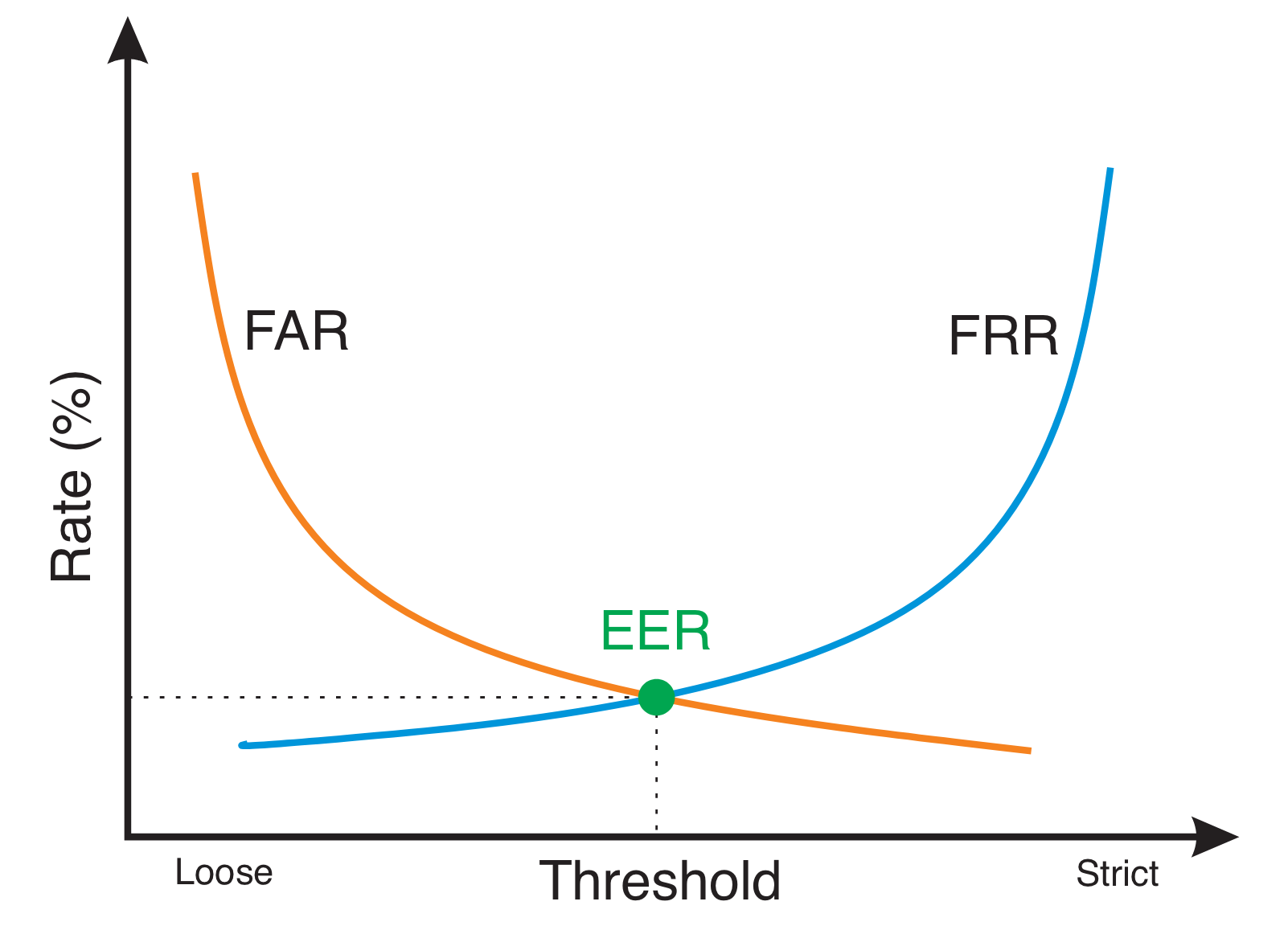}}
   \caption{Equal Error Rate (EER)}
   \end{subfigure}
\hfill
   \begin{subfigure}[b]{0.46\textwidth}
   \centering
   \fbox{\includegraphics[width=\textwidth, height=6.5cm]{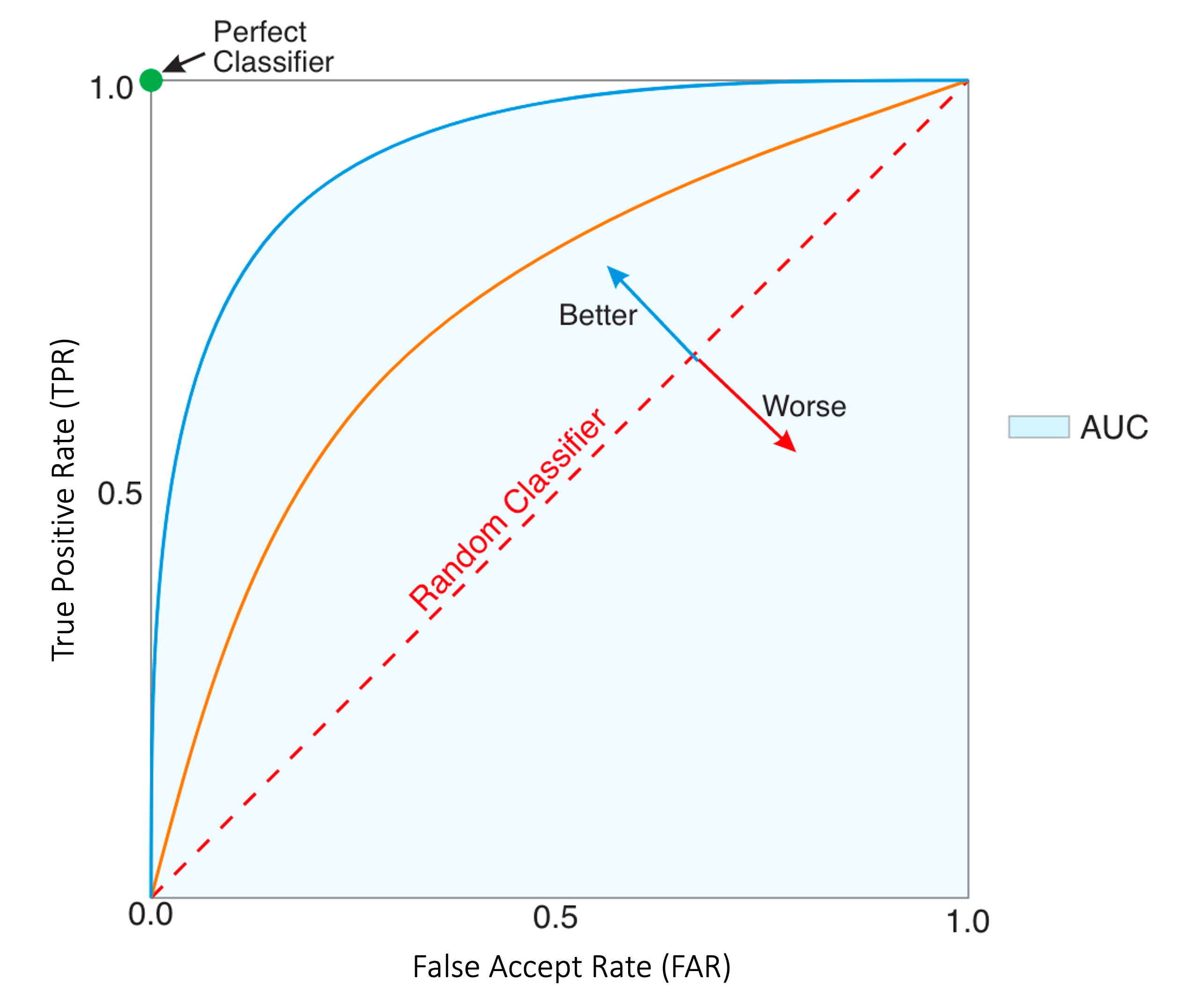}}
   \caption{Receiver Operating Characteristic (ROC) curve}
   \end{subfigure} 
\caption{(a) Equal Error Rate (EER) as the instance of False Accept Rate (FAR) and False Reject Rate (FRR) intersecting, and the effect of loose and strict threshold (when the threshold is loose, the FAR is high and FRR is low; when the threshold is strict, the FAR is low and FRR is high). The EER balances between loose and strict thresholds. (b) The Receiver Operating Characteristic (ROC) curve and the Area Under the Curve (AUC) show the performance of different classifiers. A classifier touching the point (0.0, 1.0) is desired and considered a perfect classifier. This is difficult in practical systems; a classifier closer to the coordinate (0.0, 1.0) is preferred. The ROC curve below the diagonal line is considered worse, while the ROC curve above the diagonal line is considered better. }
\label{fig:eer}
\end{figure}

The equal error rate (EER) is also a commonly used performance metric, especially when it is challenging to find an operating point between the FAR and FRR.
The EER indicates the instance of FAR and FRR intersecting. Similar to the FAR and FRR, a lower EER value is preferred. As shown in Figure \ref{fig:eer}(a), a threshold closer to 0 favors convenience (low FRR) but is less secure (high FAR), while a threshold closer to 1 favors security (low FAR) but is less convenient (high FRR). The EER balances these two points.

\subsubsection{ROC and AUC}
The Receiver Operating Characteristic (ROC) curve is a graphical plot of FAR and TPR (i.e., 1 - FRR) that visualizes the performance of a binary classifier while the threshold for classification is changed. This metric provides a clear visualization of keystroke dynamics performance across varying classification thresholds. A perfect classifier will have a point at the top left corner with coordinates (0.0, 1.0) which illustrates no false accept or false reject (Figure \ref{fig:eer}(b)). Since this is hardly practical in a large-scale system, a classifier closer to the coordinate (0.0, 1.0) is desired. The diagonal from the top right to bottom left corners indicates a random classifier, which is similar to a random guess and therefore not desired. The Area Under the Curve (AUC), often associated with the ROC curve, provides a combined performance measurement across all varying thresholds of classification.

\subsubsection{ANIA and ANGA}
In keystroke dynamics, especially in continuous user authentication, performance metrics such as FAR, FRR, ROC, and AUC may not be sufficient, as they fail to illustrate the average keystroke count required for a genuine user to be rejected or an impostor to be detected. Therefore, to further evaluate the performance of a continuous user authentication (CUA) system, the Average Number of Impostor Actions (ANIA) and Average Number of Genuine Actions (ANGA) are used, where the actions in this case are keystrokes \cite{mondal2017study}. The ANGA is the number of legitimate actions users can perform before they are wrongfully denied access, while ANIA is the number of actions impostors can perform before they are detected and locked out.
ANGA is preferred to be high (in fact, it is desired that a legitimate user should never be denied access), and ANIA as low as possible, so as to reduce the amount of actions an illegitimate user can perform before detection.

The ANIA and ANGA are mostly used in keystroke dynamics continuous authentication and can be calculated as shown in \autoref{equation:ANIA} and \autoref{equation:ANGA}, where $N$ is the number of consecutive actions a user (genuine or impostor) can finish before an authentication/identification decision is made.

\begin{equation}
\label{equation:ANIA}
ANIA = \frac{\text{$N$}}{\text{1 - FAR}}
\end{equation}

\begin{equation}
\label{equation:ANGA}
ANGA = \frac{\text{$N$}}{\text{FRR}}
\end{equation}

\subsubsection{Authentication Time}
The authentication time is an additional metric used in keystroke dynamics to describe the usability of the system. This metric is the total average time necessary for collecting keystroke data and the average of the time a system takes to permit or deny user access \cite{wahab2023usability}. While the desire to achieve a very low FAR and FRR stands, it is important to achieve this with a very short authentication time. For example, the authentication system of Zheng et al. \cite{zheng2011efficient} achieved FAR and FRR of 1.30\% but with a very long authentication time (approximately 38 minutes). Such a long authentication time would leave impostors uncaught, as most impostors require significantly less time to complete attacks. Additionally, the mouse dynamics-based authentication of Shen et al. \cite{shen2012user} had 3.33\% FAR and 2.12\% FRR with about 119 seconds of authentication time. Although these FAR and FRR are higher than those of Zhen et al., it could be considered a better system based on the lower authentication time.

\subsubsection{Rank-N}
Rank-n \cite{acien2021typenet} is an evaluation method used to measure the performance of keystroke dynamics-based authentication systems. A dataset of multiple samples from each subject is used to test the system. For each subject, one of the samples is used as a reference sample and the rest of the samples are considered as test samples. A similarity metric such as Euclidean distance is used to measure the similarity of a test sample to the reference sample. The test samples are ranked based on their similarity to the reference sample. The percentage of the inclusion of the correct test sample within the top n-ranked samples is measured by the rank-n metric. If the correct test sample is included within the top 5 ranked samples 90\% of the time, then the rank-5 accuracy is 90\%. Therefore, the rank-n value needs to be lower for better performance. For a lower rank-n value, there is a better chance that the correct test sample will be ranked higher. A rank-1 accuracy of 90\% indicates that the correct test sample is ranked 1st 90\% of the time. Rank-n evaluation provides a comprehensive analysis of the system's performance.

\vspace{4mm}
There are other performance metrics such as precision, recall, ACC (accuracy), BA (balanced accuracy), and the F1-score which are occasionally used in keystroke dynamics.

\section{Keystroke Datasets}
A keystroke dataset is required to perform a satisfactory evaluation of keystroke dynamics-based authentication algorithms. The quality, size, and nature (fixed, free, or semi-fixed text) of the dataset are crucial factors in evaluating the performance of these algorithms. To perform an evaluation, an algorithm is applied that compares the typing patterns of users based on keystroke features (see Fig. \ref{fig:Keystroke_feature}). This information is used to authenticate users.
This section details various keystroke datasets and their effectiveness in terms of authentication performance. A summary of the keystroke datatsets is shown in Table \ref{table:existingDatasets}.

\begin{table} [ht]
  \centering
  \scriptsize
  \caption{A summary of publicly available keystroke datasets. The table includes the type of data collected, the number of subjects that participated in the data collection, and the average number of keystrokes per subject.}
  \label{table:existingDatasets}
  \setlength{\tabcolsep}{5pt}
  \renewcommand{\arraystretch}{2}
  \centering
  \resizebox{\textwidth}{!}{%
  \begin{tabular}{|m{4.5cm}|m{2cm}|m{1.6cm}|m{1.6cm}|m{4cm}|m{4cm}|}  
    \hline
    		{} & {} & {} & {} & {} & {}\\[-4ex]
        
        {\textbf{Dataset}} & {\textbf{Type of data}} & {\textbf{Number of subjects}} & {\textbf{Average number of keystrokes per subject}} & {\textbf{Notes}} & {\textbf{Dataset and related software}}\\
        \hline
        {CMU (2009) by Killourhy and Maxion \cite{killourhy2009comparing}} & {Fixed} & {51} & {4400} & {Imposed password} & {\url{http://www.cs.cmu.edu/~keystroke}}\\
        \hline
        {GreyC-A (2009) by Giot, El-Abed, and Rosenberger \cite{giot2009greyc}} & {Fixed} & {133} & {Unspecified} & {Subjects practiced on other keyboards before authenticating on the actual keyboard} & {\url{http://www.ecole.ensicaen.fr/~rosenber/keystroke.html} (currently unavailable)\tablefootnote{Last attempted access: June 20, 2024.}}\\
        \hline
        {GreyC-B (2012) by Giot, El-Abed, and Rosenberger \cite{giot2012web}} & {Fixed} & {48} & {Unspecified} & {Imposed and individual credentials} & {\url{http://www.epaymentbiometrics.ensicaen.fr/} (currently unavailable)\tablefootnote{Last attempted access: June 20, 2024.}}\\
        \hline
        {Pace (2013) by Bakelman et al. \cite{bakelman2013keystroke}} & {Fixed} & {30} & {220} & {Password and numeric input} & {Unknown}\\      
        \hline
        {Clarkson I (2014) by Vural et al. \cite{vural2014shared}} & {Fixed \& Free} & {39} & {21,533} & {Includes videos of subjects' facial expression and hand movements} & {Dataset available upon request from the authors}\\
        \hline
        {OhKBIC (2015) by Monaco et al. \cite{monaco2015one}} & {Free} & {64} & {minimum 1,500} & {One-handed} & {Unknown}\\
        \hline
        {Buffalo (2016) by Sun, Ceker, and Upadhyaya \cite{sun2016shared}} & {Fixed \& Free} & {157} & {17,000} & {Desktop, and Multi-keyboard} & {\url{http://cubs.buffalo.edu/research/datasets}}\\
        \hline
        {Clarkson II (2017) by Murphy et al. \cite{murphy2017shared}} & {Free} & {103} & {125,000} & {Desktop, and Completely uncontrolled} & {Dataset available upon request from the authors}\\
        \hline
        {Aalto Desktop (2018) by Dhakal et al. \cite{dhakal2018observations}} & {Free} & {168,000} & {800} & {Desktop} & {\url{http://userinterfaces.aalto.fi/136Mkeystrokes}}\\
        \hline
        {Aalto Mobile (2019) by Palin et al. \cite{palin2019people}} & {Free} & {37,370} & {Unspecified} & {Mobile} & {\url{https://userinterfaces.aalto.fi/typing37k}}\\
        \hline
        {FACT (2020) by Kim and Kang \cite{kim2020freely}} & {Free} & {50} & {4000} & {Mobile and bilingual} & {\url{https://doi.org/10.1016/j.patcog.2020.107556}}\\
        \hline
        {AR (2021) by Wahab et al. \cite{wahab2021utilizing}} & {Semi-Fixed} & {44} & {2,048} & {Account recovery} & {Dataset available upon request from the authors}\\
        \hline      
        {Multi-K (2022) by Wahab et al. \cite{wahab2022shared}} & {Free} & {86} & {24,000} & {Multi-keyboard and bilingual} & {Dataset available upon request from the authors}\\
        \hline
  \end{tabular}}
\end{table}

\subsection{CMU Dataset (2009)}
Killourhy and Maxion \cite{killourhy2009comparing} collected fixed text (password) keystroke data and evaluated
 14 algorithms with that dataset. 51 subjects were included in the data collection. The fixed password was  
 ``\textit{.tie5Roanl}''. Each subject typed this password 400 times during 8 sessions, with 50 attempts in each 
 session. There was at least a one-day gap between successive sessions to reflect the temporal variation in 
 typing of the subjects. A Microsoft Windows application was used for the data collection. An external reference clock was utilized to make more accurate timestamps. Killourhy and Maxion evaluated different algorithms with 
 this dataset and achieved an EER between 9.6\% and 10.2\%.

\subsection{GreyC-A Dataset (2009)}
Giot, El-Abed, and Rosenberger \cite{giot2009greyc} proposed a benchmark database and software for measuring the performance of authentication systems leveraging keystroke dynamics for user access control. Their focus was on static keystroke authentication (such as passwords or passphrases). GREYC-Keystroke was the proposed software. They developed a standard
keystroke dynamics database using the GREYC-Keystroke software. Keystroke data was collected for 133 participants including researchers, students, and employees. Each participant typed the password between 5 and 107 times. The authors collected the data of 7555 attempts. Most participants attended at least 5 sessions. On average, there were 51 attempts per participant. 

Giot et al. \cite{giot2009keystroke} used the GREYC keystroke dataset to evaluate their proposed technique based on SVM learning. The enrollment step was limited to 5 captures. The administrator set a passphrase that the users had to type. A two-class SVM was utilized for the enrollment. This SVM-based machine learning approach
only needed five captured samples to develop the model. They used four types of algorithms for their experiment: 1) Statistical algorithm,
2) Distance-based algorithm, 3) Rhythm-based algorithm, and 4) Machine learning based algorithm. They also investigated the variations in the authentication method
due to different keyboards. The authors also considered the consequences of some other factors like differing amounts of samples needed to develop the model, individual or global
threshold, the database size, etc. The EER was calculated by using ten vectors from the head of the dataset for enrollment and the remaining samples were used for verification. In terms of 
EER, their proposed method performed better than all other algorithms. 

\subsection{GreyC-B Dataset (2012)}
Giot, El-Abed, and Rosenberger \cite{giot2012web} created a new dataset for keystroke-based biometric systems. To build the dataset, two types of 
logins and passwords were chosen and imposed. Leveraging a web environment, the data was collected without any restrictions. They performed
statistical analysis of some key factors including the effect of password size on performance, the significance of fusion techniques, etc. The data collection process consisted of three steps. 1) Ten attempts using a prior researcher-chosen login and password, 2) Ten attempts for the user-chosen login and password, and 3) Ten imposter samples from two 
other participants. 83 individuals participated in the data collection. They collected 5185 genuine samples, 5439 imposed samples, and 5754 imposter samples. 
For each participant, 20 samples were used for training and the remaining samples were used for testing (with a minimum of 20). The conclusions of their experiment were: 1) Using an individual threshold was better than using a global threshold for computing the
EER because an individual threshold is a value specific to a user that takes into account the origin of the sample i.e whether a legitimate user or an imposter, 2) Using logins provided better results than using passwords, 3) The performance improved by doing the fusion of all features (monographs and digraphs), 4) The performance was dependent on the password size and entropy. 

 \subsection{Pace Dataset (2013)}
Bakelman et al. \cite{bakelman2013keystroke} collected fixed-text numeric keypad data for keystroke authentication of 30 subjects over four days. On each day a maximum of 60 samples 
were collected from each subject. There were 11 keystrokes per sample from numeric sequence and the
\emph{Enter} key. The data was collected by a third-party keylogger. The subjects had to type the 
samples with only their right hand in order to simulate typing a phone number or ATM PIN. 
There were 20 samples per subject. For this dataset, they got an EER of 6.1\%.

\subsection{Clarkson I Dataset (2014)}
Vural et al. \cite{vural2014shared} developed a new dataset containing short pass-phrases, long text transcription (fixed-text) 
and free text. They also recorded videos showing subjects during data collection in order to capture the facial 
expressions and hand movements of the subjects. 
The dataset includes 2 kinds of fixed text (password and transcriptions) and free-text from 39 participants including university employees and students. 
For each participant, the data was collected in two one-hour sessions each occurring on a different day. 
The first task was typing 3 different passwords. The subjects typed each password 20 times. 

For the second task, the subjects had to answer 9 survey questions in free-text with a minimum of 500 characters for each answer. The last task was copying the commencement speech of Steve Jobs at Stanford 
University. A browser-based keylogger was used for the data collection. The data was collected for a period of 11 months between August 2011 
and June 2012 in a lab at Clarkson University. On average, each subject provided 11066 keystrokes in session 1 and 10467 keystrokes in session 2. Every session recording was analyzed to determine digraphs and trigraphs. N-graphs greater than 500 milliseconds or shorter than 30 milliseconds in length were 
discarded. In session 1, on average there were 8406 digraphs and 389 unique digraphs per subject, 7561 trigraphs, and 1603 unique trigraphs per subject. 
In session 2, on average there were 8261 digraphs and 388 unique digraphs per subject, 7520 trigraphs, and 1620 unique trigraphs per subject. 

To evaluate this dataset, 
the algorithm of Leggett et al. \cite{leggett1991dynamic} was used. A FAR of 0.25\% and FRR of 17.65\% were obtained for a threshold of 0.9; aFAR of 3.45\% and FRR of 8.82\% were obtained for a threshold of 0.85.
Ceker and Upandhyaya \cite{cceker2016user} utilized this dataset to examine the accuracy of a single-class support vector machine (SVM) on longer text data, as opposed to shorter password text. They found that using the four ranking digraphs by commonality (\textit{`he', `re', `th' and `an'}) resulted in a 2.94\% EER while using the most common 12 or more digraphs led to an almost 0\% EER.

\subsection{OhKBIC Dataset (2015)}
Monaco et al. \cite{monaco2015one} described the results of the One-handed Keystroke Biometric Identification 
Competition (OhKBIC). In the dataset, Monaco et al. included the data of normal typing and handicapped typing (typing with one hand). Data was collected 
from three online exams of undergraduate students. There were 64 subjects. In each exam, there were five essay 
questions. The subjects typed normally in the first exam. The subjects were directed to type with their left hand only for 
the second exam and with their right hand only for the third exam. Each subject typed a minimum of 500 keystrokes on 
each exam. The data was collected by a JavaScript-based keylogger and transferred to a server. Each used different keyboard make and models
throughout the three exams which was a drawback of this dataset. A portion of the data 
collected from the first exam of normal both-hands typing was used as labeled training data. 
The unlabeled data included all three types of data (both-hand) typing, left-hand typing, and right-hand 
typing data. The top team used two regression models (Artificial Neural Network and Counter-Propagation Artificial Neural Network) 
and one prediction model (Support Vector Machine) for pairwise coupling \cite{hastie1997classification}. The best 
recognition rate ranged from 55.7\% to 82.8\% for both-hand sections, 19.9\% to 30.5\% for 
the left-hand section, and 24.8\% to 34.3\% for the right-hand section.

\subsection{Buffalo Dataset (2016)}
Sun, Ceker, and Upadhyaya \cite{sun2016shared} developed a shared dataset containing keystrokes collected from 157 participants. 
The data was both fixed and free-text. The data collection process continued for 4 months 
between September and December 2015. For each subject, the data was collected in 3 sessions. The tasks included 
fixed-text transcription and answering questions in free text. On average there was a 28 days gap between 
2 sessions to consider the temporal variations. There are 2 sections in the dataset- in one section the subjects 
used the same keyboard across sessions. In the other section, the subjects used different keyboards across sessions. 
The first task was copying the 2005 commencement speech of Steve Jobs at Stanford University. The speech was 
divided into 3 parts of equal length; the subjects had to type one part in each session. The second task consisted 
of multiple subtasks including answering survey questions and describing a picture, sending an email with an attachment, 
and free internet browsing. Each session lasted for about 50 minutes with 30 minutes for the first task and 20 minutes for the
second task. A system logger was adopted for the data collection on the Windows platform 
\cite{garg2013user}\cite{sun2015secure}. Among the 157 subjects, the data of 148 subjects was included in the 
dataset. The data of the remaining users was discarded due to mistakes. On average each subject typed about
5700 keystrokes in each session, around 17000 keystrokes for all 3 sessions. There was a 3 to 5 week time gap between 
successive sessions to reflect the temporal effect. The Gaussian mixture model, a probabilistic model for classifying data into different categories based on the probability distribution, was utilized to evaluate this dataset. 
The EER was 0.01\% and 0.39\% for the Gaussian mixture model with one and two components, respectively.

\subsection{Clarkson II Dataset (2017)}
A novel dataset was proposed by Murphy et al. \cite{murphy2017shared} that included keystrokes, mouse events, 
and active programs. 103 users contributed the data for a period of around 2.5 years. Unlike 
the other datasets, the data was collected on the users' personal computers during normal use. Keystrokes and mouse data were recorded while users make use of their computers, such as for typing, gaming, etc thereby classifying the data as free-text. The users 
provided 12.9 M keystrokes, with each user providing an average of 125K keystrokes. A keylogger was installed 
on each user's computer to record keystroke and mouse movement events. The recorded data was sent to a 
remote database server. 10 samples, each sample consisting of 1000 keystrokes, were used to form the 
reference profile of a user. The rest of the user's data was considered genuine test samples. For each user, 
the data of all other users was considered as imposter data. The EER was 10.36\% for this dataset. This result 
was worse compared to fixed-text and controlled free-text datasets because it is a completely uncontrolled dataset and as such contains several keystrokes that are not usually seen in typing tasks.

\subsection{Aalto Datasets (2018; 2019)}
The Aalto University desktop \cite{dhakal2018observations} and mobile \cite{palin2019people} datasets are large-scale datasets collected using an online typing test on desktop computers and mobile devices. The desktop dataset has 136 million keystrokes gathered over three months from 168,000 subjects. Participants were instructed to transcribe fifteen sentences in English by typing them as fast and accurately as possible. The 15 sentences were randomly taken from a set of 1,525 samples consisting of at least 3 words and a maximum of 70 characters.
Subjects were allowed to make typing errors, correct them, or add new characters when typing, and as a result, they could type more than 70 characters.
The dataset is categorized as a controlled free-text dataset because it involves transcribing, subjects did not type contents of their own but were shown what to type. There were 15 sessions and participants transcribed only one sentence during each session.
The front end of the online test was deployed as an HTML web page leveraging CSS for website styling. JavaScript was used for dynamic events on the website such as keypresses and the data were stored in a MySQL database. 

The Aalto mobile dataset \cite{palin2019people} was collected from 37,370 participants performing transcription services on a website.
It is an extension of the Aalto desktop dataset and follows the same procedure but on mobile devices. 
Subjects were to memorize a given sentence, and then type it quickly and accurately. 
Being a web-based method with a browser-side logger, as opposed to laboratory-based, there was less control over the quality of the data collected, and web applications on mobile devices have limited access privileges. As a result, many participants' data were found with undefined keycodes; many devices' down and up keystroke events were equal, producing invalid keystroke duration of $<$10ms; and the keycodes of some pressed keys were missing.

\subsection{AR Dataset (2021)}
Wahab et al. \cite{wahab2021utilizing} collected a new dataset of 500,000 keystrokes, designed for account recovery, to determine the performance of five 
differing keystroke dynamics-based algorithms \cite{gunetti2005keystroke}, \cite{huang2017benchmarking}, \cite{killourhy2009comparing}, \cite{killourhy2010did}: Euclidean distance, Manhattan distance, Scaled Manhattan distance, Mahalanobis distance, and the Gunetti and Picardi's algorithm. 
A total of 44 subjects were used during data collection including both students 
and university employees. The subjects had to complete an account recovery form that 
contained different fields. This type of data is called semi-fixed as it is neither 
free-text nor fixed-text. Two collection sessions were hosted to gather the data. In the first session, the 
subjects had to fill out an enrollment form ten times. The data collected from session one 
was used to create the profile of a subject. In the second session, the subjects had to fill out 
the same form again five times. The data collected from session two was used as genuine keystroke data 
for the corresponding subject and imposter data for other subjects. There was a one or two-week gap 
between the sessions. The fields of the enrollment form were- \emph{Full name}, \emph{Address}, 
\emph{City}, \emph{Zip}, \emph{Phone}, \emph{Email}, \emph{Declaration}, \emph{Password}. 
28 subjects attended the second session and 16 subjects 
acted as imposter. Five algorithms were used for the evaluation of this account recovery dataset. 
The Scaled Manhattan distance algorithm showed the best performance. The best EER was 5.47\% for 
individual fields, 0\% for five fields combined, and 0\% for seven fields combined. 

\subsection{Multi-K dataset (2022)}
Wahab et al. \cite{wahab2022shared} created two novel free-text keystroke dynamics datasets. The first was a multi-keyboard keystroke dataset.
Data was collected by leveraging four physical keyboards: mechanical, membrane, ergonomic, and laptop keyboards. For the second dataset, data was collected in both English and Chinese languages. 86 users participated in the data collection: 60 users for the multi-keyboard dataset
and 26 users for the bilingual dataset. A keylogger deployed on a website was leveraged for the collection of both datasets.

To examine the impact of multi-keyboard and multi-language data in keystroke dynamics, Wahab et al. employed two free-text algorithms: the Instance-based Tail Area Density (ITAD) metric and the D-Vectors model. On average, each user provided around 14,000 keystrokes for the multi-keyboard dataset. For each user, per keyboard type, data from the initial visit was used for enrollment, while data from the second visit was used for testing. There were 200 keystrokes in each test sample, and the performance was measured using the Equal Error Rate (EER). The results indicated that enrolling and testing with different keyboards affect keystroke dynamics performance where keyboard size and layout were two significant factors.

In the bilingual experiment, their aim was to investigate the impact of multi-language data in keystroke dynamics. On average, each user provided around 10,000 keystrokes for both English and Chinese languages. Monograph and digraph features were extracted from the dataset. Data from the first two questions was used for enrollment and data from the other questions was used for testing, with around 900 keystrokes used for enrollment and each test sample consisting of 200 keystrokes. The results revealed a performance degradation for cross-language in keystroke dynamics, with a performance loss of 14\% when enrollment was done with Chinese data and testing was done with English data, and a loss of 6.4\% when enrollment was done with English data and testing was done with Chinese data.

\subsection{Touch Screen and Mobile Datasets}
El-Abed, Dafer, and El Khayat \cite{el2014rhu} developed a new benchmark named RHU Keystroke for touch screen keystroke dynamics. They used a
Windows Phone app for data collection. The data was stored in a keystroke dynamics database and necessary features were extracted.
The participants in the data collection were students of varying backgrounds to ensure the scenario was as real as
possible. A total of 53 participants participated in 3 sessions typing the password \emph{rhu.university} 15 times in each session. On average there was a 5 days gap between successive sessions. The RHU Keystroke benchmark consists of four timing features: PP, PR, RP, and RR.

Stragapede et al. \cite{stragapede2023behavepassdb} created a new database of mobile Human-Computer Interaction (HCI) named BehavePassDB. 81 users participated in the data collection sessions through a dedicated mobile app installed on the user's device. The user was asked to perform a series of tasks. In order to check whether the model can differentiate between the notion of user and device, two dedicated sessions were carried out by another person (imposter) on the same exact device at the time of data collection. Data was collected from 15 background sensors as well, i.e., touchscreen, accelerometer, etc. For evaluation, Long-Short Term Memory (LSTM) architecture was used. 

Al-Obaidi and Al-Jarrah \cite{al2016statistical} proposed a classifier leveraging a median-based method for keystroke authentication on mobile devices. The classifier
was used as an anomaly detector on the SU dataset \cite{antal2015keystroke}. For the SU dataset, the CMU password (``.tie5Roanl'') 
was used. There were both timing and touch features of mobile devices in this dataset. There were 11 characters in the password including the 
Enter key resulting in 71 features in total (timing and touch screen features). 42 users took part in the data collection typing the same 
password 51 times. 34 entries were used for training and 17 for testing. Data collection was performed on Android devices, a tablet, and 
a mobile phone. The proposed statistical classifier model used the median of the training features. The testing features were authenticated against the 
training features. There were three timing features: Hold (H), Up-Down (UD), Down-Down (DD), and two touch screen features: Pressure (P) and Finger Area
(FA). The first five entries of each subject were used to create an imposter set. Al-Obaidi and Al-Jarrah achieved an EER of 8.53\% for 41 timing features and an EER of 6.79\%
for both timing and touch screen features. 

Al-Obaidi and Al-Jarrah \cite{alnoor2016statistical} proposed a keystroke authentication system for use on mobile devices. They adopted an anomaly detector leveraging statistical distance-to-median that used both the timing and touch screen features. The goal was to develop a suitable keystroke dataset for mobile devices and evaluate the
performance of the distance-to-median anomaly detector. They performed their experiment on Nexus smartphones and tablets and used five features for 
the authentication system: Hold, Latency or Up-Down (UD), Down-Down (DD), Pressure (P), and Finger Area (FA). The system consisted of two phases. 
In the training module, the required tasks were: 1) Registration of users through user-id and password, 2) Selection of the number of attempts of password entry and
the pass-mark, 3) Password typing for the selected number of times, and 4) Template generation. In the testing phase, the test vector was compared against the template stored 
in the database. The test vector was generated from the login attempt. The data was collected from 56 users each typing a 10-character password 51 times. 
For a different pass mark per subject, they achieved an EER of 0.049. For a global pass-mark, Al-Obaidi and Al-Jarrah achieved an EER of 0.054 and implemented three verification models based on distance to obtain their results. They obtained an FRR (false rejection rate) of 5.6\% for the FAR (false acceptance rate) of 5\%. 

Kim and Kang \cite{kim2020freely} proposed a novel keystroke dynamics-based authentication method for mobile devices named FACT (free text, accelerator, coordinate, and time). In order to evaluate their proposed model, Kim and Kang collected a novel keystroke dataset based on mobile devices. They used an Android-based application for the data collection. A total of 50 subjects
participated in the data collection. 20 free-text scripts were typed by each subject and each script contained about 200 keystrokes. Special characters were not included in the scripts.
The subjects provided the data in both English and Korean. 17 features were extracted from the raw data including not only the typing pattern but also hand size, finger length, fingertip size,
and muscle flexibility. The proposed FACT method achieved an EER lower than 1\%. Kim and Kang made their collected keystroke dataset public \cite{kim2020freely}.

\section{Keystroke Authentication Algorithms}
The authentication algorithm is a crucial part of any keystroke-based authentication system.
This section describes state-of-the-art keystroke authentication algorithms. A summary of keystroke authentication algorithms is shown in Table \ref{table:authenticationAlgorithms}.

 \begin{table}[t]
  \centering
  \scriptsize
  \caption{A summary of keystroke authentication algorithms, the datasets that were used to evaluate the algorithms, and the results obtained using these algorithms.}
  \label{table:authenticationAlgorithms}
    \setlength{\tabcolsep}{5pt}
  \renewcommand{\arraystretch}{2}
  \centering
  \resizebox{\textwidth}{!}{%
  \begin{tabular}{|m{6cm}|m{4cm}|m{4cm}|}  
    \hline
    		{} & {} & {}\\[-4ex]
        {\textbf{Algorithm}} & {\textbf{Dataset}} & {\textbf{Result}}\\
        \hline
        {Distance Metric Fusion \cite{migdal2018analysis}} & {Murphy et al. \cite{murphy2017shared}} & {EER-3.6\%}\\
        \hline
        {Gaussian Mixture Model \cite{ceker2015enhanced}} & {Vural et al. \cite{vural2014shared}} & {EER-0.08\%}\\
        \hline
        {Sorted Time Mapping Technique along with Neural Networks \cite{ahmed2013biometric}} & {Ahmed and Traore \cite{ahmed2013biometric}} & {EER-2.46\%}\\
        \hline
        {Instance-Based Tail Area Density Metric combined with Fused Matching Score \cite{ayotte2020fast}} & {Murphy et al. \cite{murphy2017shared} and Sun, Ceker, and Upadhyaya \cite{sun2016shared}} & {EER-7.8\% and 3\%}\\
        \hline
        {Novel Statistical Model \cite{foresi2019user}} & {Foresi and Samavi \cite{foresi2019user}} & {FAR-0.0\% and FRR-2.54\% \& 2.87\%}\\
        \hline
        {Combination of MLP and Trust Algorithm \cite{popovici2014combined}} & {Popovici et al. \cite{popovici2014combined}} & {Classification rate-80\%}\\
        \hline
        {Transfer Learning Techniques \cite{cceker2017transfer}} & {Vural et al. \cite{vural2014shared}} & {EER-19.47\%}\\
        \hline
        {XGBoost \cite{singh2020analysis}} & {Killourhy and Maxion \cite{killourhy2009comparing}} & {Accuracy-93.60\%}\\
        \hline
        {Event Sequences \cite{syed2014leveraging}} & {Syed, Banerjee, and Cukic \cite{syed2014leveraging}} & {Increased Effectiveness}\\
        \hline
        {ECM-ELM Classification Model \cite{ravindran2015keystroke}} & {Killourhy and Maxion \cite{killourhy2009comparing}} & {Stable and Good Accuracy(86.97\%)}\\
        \hline
        {Authentication without Template \cite{singh2018keystroke}} & {Singh et al. \cite{singh2018keystroke}} & {SFR \& SFA - 87}\\
        \hline
        {Triboelectric Device \cite{wu2018keystroke}} & {Wu et al. \cite{wu2018keystroke}} & {EER-1.15\%}\\
        \hline
        {Recurrent Neural Network \cite{kobojek2016application}} & {Killourhy and Maxion \cite{killourhy2009keystroke}} & {EER-0.136}\\
        \hline 
        {mRMR Feature Selection Technique \cite{krishnamoorthy2018identification}} & {Krishnamoorthy et al. \cite{krishnamoorthy2018identification}} & {Accuracy-0.9740 and F1 score-0.9701}\\
        \hline  
        {Nearest Neighbor(New Distance Metric) and Outlier Removal \cite{zhong2012keystroke}} & {Killourhy and Maxion \cite{killourhy2009comparing}} & {EER-0.084 and ZMFAR-0.405}\\
        \hline  
        {TypeNet \cite{acien2021typenet}} & {Acien et al. \cite{acien2021typenet}} & {EER-2.2\%}\\
        \hline
        {TypeFormer \cite{stragapede2022typeformer}} & {Aalto mobile \cite{palin2019people}} & {EER-3.25\%}\\  
        \hline
        {DoubleStrokeNet \cite{neacsu2023doublestrokenet}} & {Aalto desktop \cite{dhakal2018observations} and mobile \cite{palin2019people}} & {EER-0.75\% and 2.35\%}\\
        \hline
  \end{tabular}}
\end{table}

\subsection{Statistical Algorithms}
Statistical algorithms are easy to implement, model, and explain. They are deterministic and also do not require heavy computing power and/or memory. However, they require manual extraction of features which could be difficult. 
Below, we describe common statistical algorithms used for keystroke dynamics. 

\subsubsection{Gaussian Mixture Model}
Gaussian Mixture Model (GMM) is a statistical method based on the weighted sum of probability density functions of multiple Gaussian distributions \cite{vural2014shared, deng2013keystroke}. 
GMM generates a vector of mean values corresponding to each component and a matrix of covariance which includes components’ variances and the co-variances between each other. The parameter set $\lambda$ is used for expressing GMM. The parameter $\lambda$ comprises of component weights $w\textsubscript{$i$}$, mean vector $\overrightarrow{\mu\textsubscript{$i$}}$ and covariance matrix $\Sigma\textsubscript{$i$}$ as shown in equation $\lambda = \{w_i, \overrightarrow{\mu_i}, \Sigma_i\}, \text{$i = 1, ..., M$}$. The parameters are estimated using the iterative expectation–maximization (EM) algorithm \cite{dempster1977maximum}. In every iteration, the parameter ($\lambda$) is updated if the iteration yields a higher likelihood and fits the distribution of the training dataset.

Given a set of reference samples $\lambda$ (also known as profile), the mixture density for the keystroke input sample $\overrightarrow{x}$ is defined as the weighted linear combination of M pure Gaussian distributions as shown in the equation: $p(\overrightarrow{x}|\lambda) = \sum_{i=1}^{M} w_i * p_i(\overrightarrow{x})$; where $p_i(\overrightarrow{x}) = \frac{1}{(2\pi)^{D/2}|\Sigma_i|^{1/2}}exp\biggl\{-\frac{1}{2}(\overrightarrow{x}-\overrightarrow{\mu})^T\Sigma_i^{-1}(\overrightarrow{x}-\overrightarrow{\mu})\biggl\}$.

\begin{figure}[t]
\begin{center}
   \fbox{\includegraphics[width=0.6\linewidth]{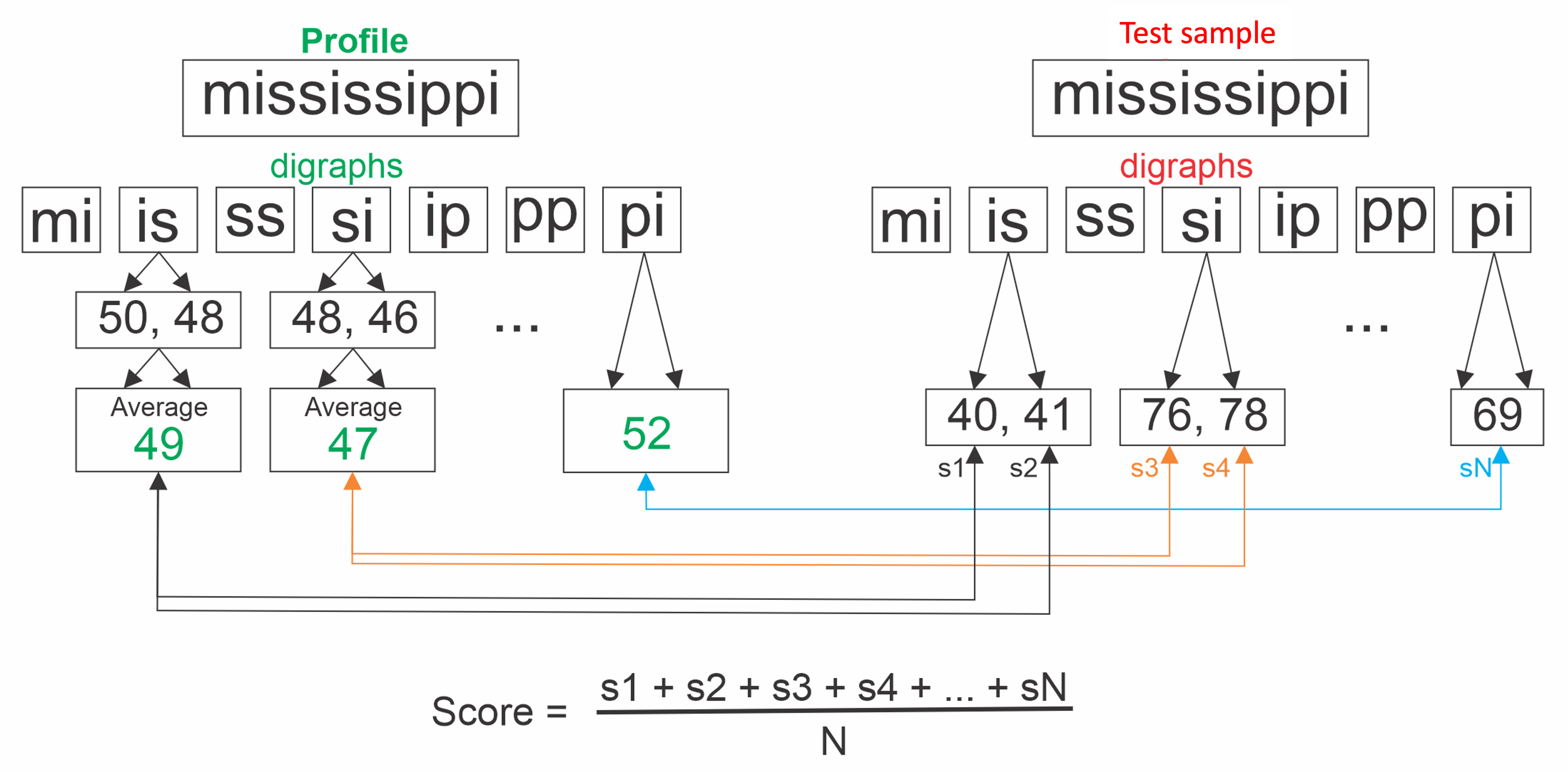}}
\end{center}
   \caption{Instance-based scoring procedure for sample text `mississippi'. The graphs in the test sample are compared with those of the profile sample and similarity scores are calculated. The
   average of the scores is returned as the final score.}
\label{procedure}
\end{figure}

\subsubsection{Euclidean Distance}
Euclidean distance is the shortest distance between two locations (points) in Euclidean space and is found in equation $D = \frac{1}{N}\sqrt{\sum_{i=1}^{N} (\mu\textsubscript{g\textsubscript{i}}-x_{i})^2}$; where $N$ represents the number of digraphs shared between the test sample and the profile, $x_{i}$ is the individual test graph duration for the $i^{\text{th}}$ shared graph in the test sample, and $\mu\textsubscript{g\textsubscript{i}}$  is the mean of the $i^{\text{th}}$ graph in the profile \cite{killourhy2009comparing, teh2013survey}. The Euclidean distance is an instance-based similarity metric and comparison is done on a single occurrence of a graph from the test sample to the reference profile as shown in Figure \ref{procedure}. The Euclidean distance assumes that the distribution of the graphs to be compared has the same variance. Hence, for two graphs with significant differences in variance, the Euclidean distance result is less accurate. Furthermore, if a correlation is present between the features, which is generally the case in real-world datasets, the Euclidean distance between a point and the mean of the points can give less accurate or misleading data about the distance between points. A reason for this is that Euclidean distance is a distance between two points only and fails to consider the relationship between a point and other points.

\subsubsection{Manhattan and Scaled Manhattan Distance}
The Manhattan distance, also known as the city-block distance, has been commonly used by keystroke researchers \cite{manhattan, tey2013can, killourhy2010did} and can be calculated in equation $D =  \frac{1}{N}\sum_{i=1}^{N} \|\mu\textsubscript{g\textsubscript{i}}-x\textsubscript{i}\|$.
$N$ is the number of digraphs shared between the test sample and the profile, $x_{i}$ is the individual test graph duration for the $i^{\text{th}}$ shared graph in the test sample, and $\mu\textsubscript{g\textsubscript{i}}$ is the mean of the $i^{\text{th}}$ graph in the profile. The Manhattan distance also follows an instance-based procedure (Figure \ref{procedure}). Similar to the Euclidean distance, the Manhattan distance assumes that the test and profile graphs have the same variance, which is often not the case. Hence, the Scaled Manhattan distance, a modified version, is used. It can be calculated using equation $D = \frac{1}{N}\sum_{i=1}^{N} \frac{\|\mu\textsubscript{g\textsubscript{i}}-x\textsubscript{i}\|}{\sigma\textsubscript{g\textsubscript{i}}}$. The Scaled Manhattan distance is similar to the Manhattan distance, except that the former is divided by the standard deviation ($\sigma\textsubscript{g\textsubscript{i}}$) of the $i^{\text{th}}$ graph in the profile \cite{killourhy2009comparing}.

\begin{figure}[t]
\centering
   \begin{subfigure}[b]{0.46\textwidth}
   \centering
   \fbox{\includegraphics[width=\textwidth]{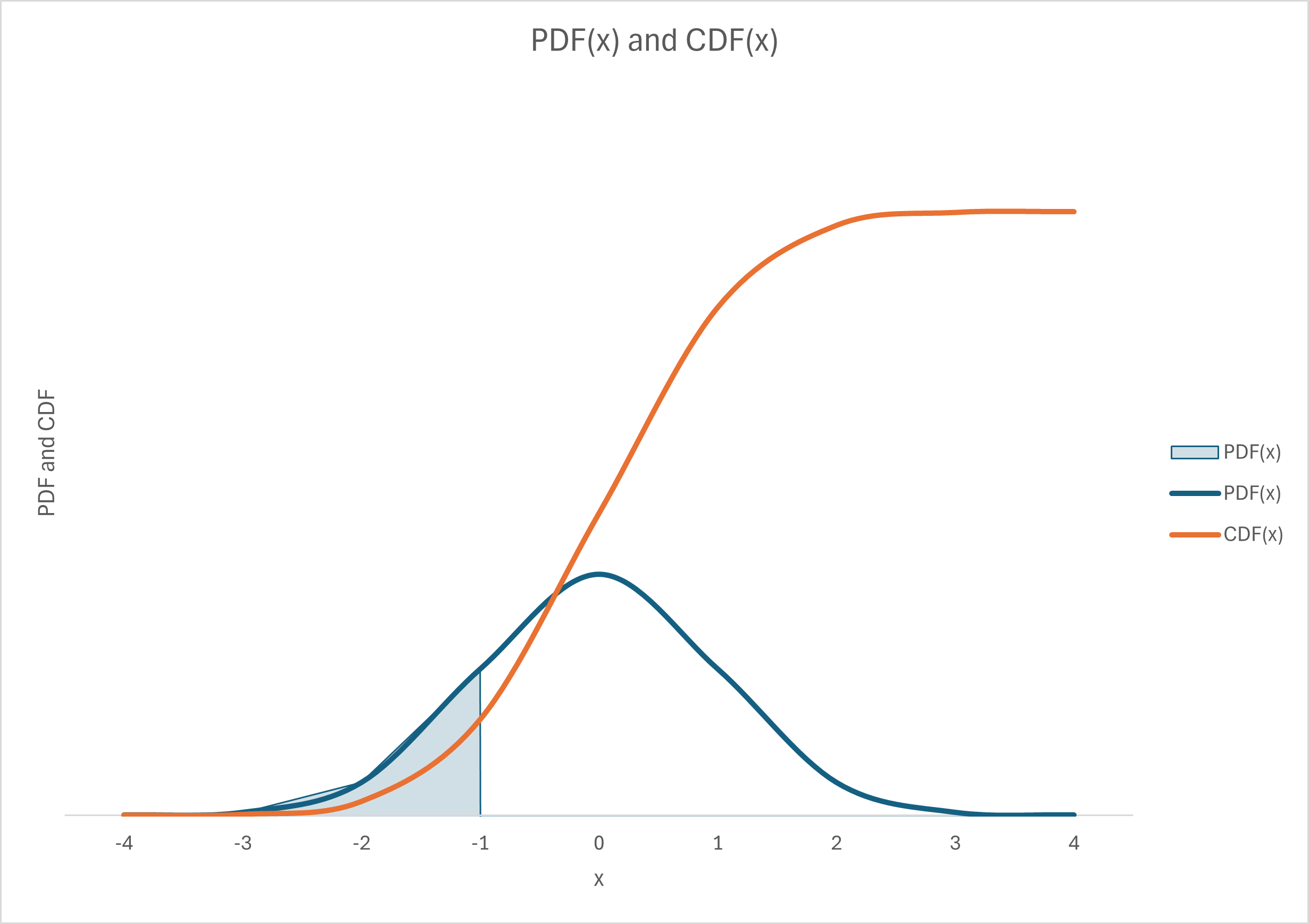}}
   \caption{Sample below the median}
   \end{subfigure}
\hfill
   \begin{subfigure}[b]{0.46\textwidth}
   \centering
   \fbox{\includegraphics[width=\textwidth]{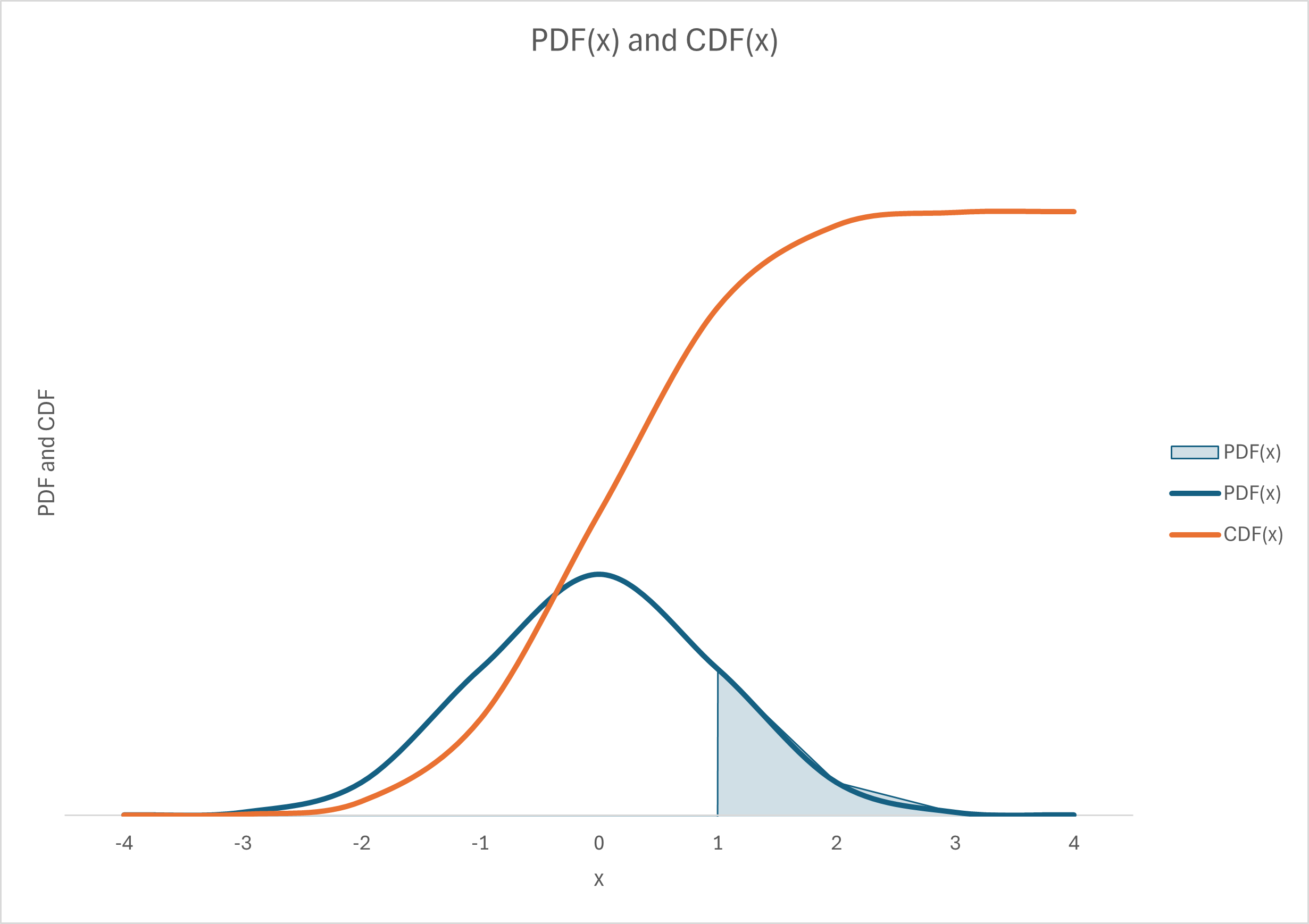}}
   \caption{Sample above the median}
   \end{subfigure} 
\caption{Graphical representation of how the ITAD metric is computed from the PDF and CDF. In (a), the sample is below the median and the ITAD metric is determined by the left tail area of the PDF. If the left tail area is x, then the x point of the CDF (on the left) indicates the ITAD metric score. In (b), the sample is above the median and the ITAD metric is determined by the right tail area of the PDF. If the right tail area is y, then the (1-y) point of the CDF (on the right) indicates the ITAD metric score.}
\label{fig:itad}
\end{figure}

\subsubsection{Instance-Based Tail Area Density}
Ayotte et al. \cite{ayotte2020fast} proposed the instance-based tail area density (ITAD) metric, a novel instance-based graph comparison algorithm, to decrease the required number of keystrokes for authentication. The ITAD metric is also an instance-based similarity metric (Figure \ref{procedure}). This metric uses the tail area under the probability density function (PDF) or the percentile value of each test graph in the profile sample. Alternative to the instance-based algorithm is the distribution-based algorithm \cite{ayotte2020fast}. Distribution-based algorithms compare the PDF of each graph in the test and profile samples.
Unlike distribution-based which requires that there are at least 4 instances of a graph in the test and profile samples before they can be compared, instance-based algorithms (such as ITAD) can compare graphs in the test and profile samples even when there is just a single instance of that graph in the test sample.
ITAD can be calculated as shown in Equation \ref{area_eqn},
\noindent where CDF\textsubscript{g$i$} is the empirical cumulative distribution function of the $i$th graph in the gallery (profile), M\textsubscript{g$i$} is the median of the $i$th graph in the gallery, x\textsubscript{$i$} is the individual test graph duration for the $i$th graph in the test sample that was shared with the gallery, and $N$ is the number of graphs shared between the test sample and the gallery. The ITAD metric scores for N graphs are averaged and a single similarity score is returned as given in Equation \ref{area_eqn2}.

The illustration in Figure \ref{fig:itad} depicts the computation of the ITAD metric using a graphical representation. The ITAD metric is calculated as the area of the tail of the PDF. If the sample falls below the median, the ITAD metric is determined by the tail area on the left, while if it is above the median, it's determined by the tail area on the right.
ITAD performs better than any distance metrics (such as the Scaled Manhattan) which are based on mean and standard deviation of graphs and are affected by outliers. ITAD metric scores range between 0 and 0.5 and can be interpreted as a measure of similarity. The larger the score is, the closer the test sample is to the profile.

\begin{equation}
    S_\text{$i$} = 
\begin{dcases}
    CDF_\text{g$i$}(x_\text{$i$}),& \text{if } x_\text{$i$}\leq M_\text{g$i$}\\    
    1-CDF_\text{g$i$}(x_\text{$i$}),& \text{if } x_\text{$i$} > M_\text{g$i$}
\end{dcases}
\label{area_eqn}
\end{equation}
\begin{equation}
\label{area_eqn2}
\text{Similarity Score} = \frac{1}{N}\sum_{i=1}^{N} S_\text{$i$}
\end{equation}

\subsubsection{Mahalanobis Distance}
Mahalanobis distance is an example of a distribution-based algorithm. It is an effective multivariate distance metric that measures the distance between a point and a distribution \cite{sae2022distinguishability, zhong2012keystroke, zhong2015survey, killourhy2009comparing}. It takes into consideration the correlation between features which is more suitable when features are not independent. It is best for finding multivariate outliers and finds its applications in multivariate anomaly detection, classification, and one-class classification. The Mahalanobis equation is: $D = (X - \mu_t)\cdot C^{-1}\cdot (X - \mu_t)$; where $X$ is the feature vector of a test sample, $\mu_t$ is the mean feature calculated from the profile template, and $C^{-1}$ is the inverse covariance matrix of the user profile template.

\subsubsection{\textbf{Gunetti and Picardi's Metric}}
Gunetti and Picardi's free-text algorithm~\cite{gunetti2005keystroke} combines typing speed (A-measure) and the degree of disorder (R-measure) to measure similarity. 
The `A' measure represents the distance between typing samples $S_1$ and $S_2$ in terms of $n$-graphs, where $n$ is the number of consecutive keystrokes ($n=1$ is monograph, $n$=2 is digraph, $n$=3 is trigraph, etc.),  is given in equation - 
\begin{equation}
A\textsubscript{t,n}(S_1,S_2) = 1 - \frac{\text{Number of similar n-graphs between S\textsubscript{1} and S\textsubscript{2}}}{\text{Total number of n-graphs shared by S\textsubscript{1} and S\textsubscript{2}}}
\end{equation}
where \textit{t} is a constant for determining n-graph similarity. 
For example, let G\textsubscript{$S_1$,$L_1$} and G\textsubscript{$S_2$,$L_2$} be the same n-graph occurring in typing samples $S_1$ and $S_2$, with duration $L_1$ and $L_2$, respectively. 
We say that G\textsubscript{$S_1$,$L_1$} and G\textsubscript{$S_2$,$L_2$} are similar if and only if 1 $\le$ max($L_1$,$L_2$)/min($L_1$,$L_2$) $\le$ \textit{t}.

The `R' measure on the other hand quantifies the degree of disorder between two sequences. Given an array $V$ of $K$ elements, a simple measure of the degree of disorder of $V$ with respect to $V^\prime$ (its ordered counterpart) is computed as the sum of the distances between the position of each element in $V$ and the position of the same element in $V^\prime$.

\subsection{Machine Learning-Based Algorithms}
Machine learning methods have been increasingly employed for keystroke dynamics. Here, we will detail the most commonly used machine learning-based methods that are used for keystroke dynamics.

\subsubsection{Supervised Learning}
Supervised Machine Learning results in a model that predicts the outcome/label of a given data after the model has been trained with labeled data. The model at training has access to the label of the input (training) data which is used for computing training errors through the loss function. Based on training errors, the model backpropagates and updates the weights. When model training is complete, it is used to predict unseen input data in the absence of labels. Model accuracy is represented by the number of accurate predictions over the total number of predictions.

Supervised Machine Learning can be further divided into classification and regression. Classification uses an algorithm to accurately predict a class label on test data e.g., classifying user behavior as genuine or imposter. Binary classification refers to a problem involving two classes and multi-class classification refers to a problem with greater than three classes. On the other hand, regression makes \emph{numerical} predictions such as predicting the weather or average salary. In keystroke dynamics, the focus is either to verify if the user is who he/she claimed to be (authentication) or to identify the user to who the query sample belongs to (identification). Both scenarios belong to the classification type, hence, keystroke dynamics uses the classification techniques.  

\paragraph{Support Vector Machine}
A support vector machine (SVM) is a type of supervised learning model leveraged for classification problems. 
The SVM algorithm computes hyperplane(s) that uniquely separate classes of points of data. The hyperplane, also known as the decision boundary, is an N-dimensional space plane, where N is the number of features in the dataset. Of several hyperplanes possible, the goal is to compute a boundary with the maximum margin that separates the classes (legitimate users from impostors) \cite{giot2009keystroke, krishnamoorthy2018identification, wu2018keystroke, singh2020analysis}. This algorithm derives its name from support vectors, which are data points that are closer to the hyperplane and decide the orientation of the plane.

In keystroke dynamics, it is assumed that the keystroke samples in the instance of enrollment and verification have an equal distribution, which is not usually the case, and as a result, lower prediction accuracy is obtained from most traditional machine learning classifiers. In this regard, SVM has support for adaptive learning \cite{cceker2017transfer}, where an existing SVM classifier (source) is adapted to another SVM classifier (target) based on some projection techniques. Knowing that there exist variations in users' typing patterns at every sample, especially when the samples have a wide time margin (weeks, months, or years apart), 
the adaptive SVM modifies the hyperplane of a pre-trained classifier using the new data points with minimum retraining and therefore performs better than the regular SVM trained on the same amount of labeled data. The SVM classifier is a commonly used classifier in keystroke dynamics, especially on a small feature set dataset \cite{cceker2016user, giot2011unconstrained}.

\paragraph{Naive Bayes}
The Naive Bayes classifier is a probabilistic machine learning model used for classification and based on the Bayes theorem with ``naive'' independent assumptions between features \cite{anusas2019strengthening, ho2018one, darabseh2015effective}. The Bayes’ theorem finds the probability of an event $a$ occurring given that the probability of another event $B$ has already occurred, as shown in equation $P(a|B) = \frac{P(B|a) \times P(a)}{P(B)}$; where $P(a|B)$ is the posterior probability, $P(a)$ is the prior probability of the label class, $P(B)$ is the prior probability of the predictor and $P(B|a)$ is the likelihood. The Naive Bayes classifier offers several advantages, such as being less computationally demanding compared to other traditional machine learning models, and performing well on small training data when the assumption of feature independence holds. Additionally, it can be used for solving multi-class prediction problems, which are identification problems that involve predicting one of several possible classes (or outcomes) for a given input. In keystroke dynamics, where dependencies between features may exist, the Naive Bayes classifier may not be as effective and is therefore less commonly used.

\paragraph{Tree-Based Models}
Tree-based models are a type of supervised machine learning algorithm used for classification. They use a set of conditional statements to recursively split training data into subsets. The end result model is a roadmap of logical decisions that describes the dataset. Tree-based models are easy to implement and interpret, less computationally intensive, and are popularly used in keystroke dynamics because of their non-parametric attributes, making it preferable even for non-Gaussian distributed data \cite{sheng2005parallel}. Tree-based models are tolerant to outliers and intra-class variation problems, work with both numerical and categorical variables, and require fewer data preprocessing steps such as variable transformations (scaling, normalization, etc). However, they are prone to overfitting and high variance. Popular tree-based methods used in keystroke dynamics are random forest \cite{ayotte2020fast, alshanketi2016improving}, gradient boost and extreme gradient boost (XGBoost) \cite{singh2020analysis, chang2022machine}

\paragraph{K-Nearest Neighbors}
K-Nearest Neighbors (KNN) is a supervised machine learning classifier (although it can also be unsupervised) that classifies new data points based on their proximity with data points of already known classes. KNN works on the assumption that data points of similar classes are closer to each other and can be found near each other \cite{hu2008k, stanciu2016effectiveness, anusas2019strengthening}.
During testing the distances between a new point of data and prior points (neighbors) are calculated. A class label is assigned to a new data point based on a majority vote between its K nearest neighbors, where K is the number of neighbors that will be checked to determine the class to assign for the new query point. Generally, the optimal value of K is found experimentally, as a low value of K results in low bias but high variance, while a high value of K gives high bias and low variance. Distance metrics (such as the Euclidean distance, Manhattan distance, Hamming distance, etc.) are used to calculate the distance between the query point and other data points. These distance metrics form the decision boundaries. KNN is known as a lazy learner as it does the nearest neighbor calculation at every run time. Because of this, it is used only on low-dimensional data and has a high computational cost when used on high-dimensional data. 

\paragraph{Multi-Layer Perceptron}
Multi-layer Perceptron (MLP) is a type of supervised learning algorithm consisting of input, output, and several hidden layers (those in between the input and output layer). The input data to be processed is passed to the input layer, the output layer does the classification, and the hidden layers (non-linear layers) perform the computational work. Therefore, data flows in the forward direction from the input layer to the output layer. MLP is a neural network with multiple layers and each layer is a combination of neurons such that the output of some neurons is passed as input to other neurons in the next layer. The number of neurons and hidden layers used is dependent on the type of classification problem being solved. MLP is capable of learning non-linear models, but it requires several hyperparameter tuning such as the number of hidden neurons, hidden layers, and iterations. Hence, it is computationally intensive and requires both genuine and imposter data to train.
The neural network is capable of producing much better performance than statistical algorithms and any other machine learning algorithms when working with big data that has both genuine and impostor samples \cite{ahmed2013biometric, harun2010performance, stylios2021bioprivacy, popovici2014combined}.

\subsubsection{Unsupervised Learning}
Unsupervised learning is a type of ML used for identifying patterns in datasets that contain unlabeled data points whereas with supervised learning the data is labeled and the algorithm is trained on this data to make predictions. In keystroke dynamics, unsupervised learning algorithms are often used for outlier detection. Some of the unsupervised learning algorithms used in keystroke dynamics include One-class SVM (OCC), K-NN, and K-Means \cite{ivannikova2017anomaly, wahab2021utilizing}. These algorithms are useful in situations where obtaining labeled data is difficult, costly, or impossible, and the ground truth (label) is unknown.

\subsection{Deep Learning-Based Algorithms}
Deep learning algorithms have become the new state-of-the-art in the field of keystroke dynamics. Deep learning is a machine learning technique that learns features automatically from data. The data in this case are sequential keystroke data. Deep learning, when modeled properly, is robust to natural variations in sequential or temporal data. In keystroke dynamics research, usually a conventional machine learning algorithm works as a binary classifier whereas a deep learning algorithm works as a global model that generates embeddings for all subjects.

Deep learning is data-hungry and its performance is partially directly proportional to the volume of the training data used, and as such, its application in keystroke dynamics has been limited due to the lack of large publicly available datasets. However, with the availability of large keystroke datasets such as the Aalto \cite{dhakal2018observations, palin2019people} and Clarkson II \cite{huang2017benchmarking} dataset, this option is now being explored \cite{acien2021typenet, acien2019keystroke}. Popular deep neural network types are the convolutional neural network (CNN) and recurrent neural network (RNN).

\subsubsection{Convolutional Neural Network}
Convolutional neural networks (CNNs) \cite{gu2018recent} are deep learning algorithms that take in an input image and learn to identify key features in the image much like human vision.
The basic CNN architecture is comprised of several layers such as the input layer, the convolution layer where high-level features are extracted from the input image, the pooling layer where the spatial size of the convolved feature is reduced, and the fully connected or dense layer where the result is flattened out and passed through an activation function that outputs the final classification result \cite{fukushima1982neocognitron}.
Basically, CNN uses convolution, also known as a sliding window (``filter”) that passes over the image, extracting key features and feeding them to lower layers in the network. A CNN, when trained with enough data, can successfully capture the spatial and temporal information in the data. Although CNNs were originally created for image classification where pixels form a two-dimensional grid, it has found its usefulness in keystroke dynamics \cite{cceker2017sensitivity, lin2018neural, xiaofeng2019continuous} and several other fields.

To leverage the CNN for keystroke analysis, keystroke data are grouped into sequences, with each sequence containing a fixed length ($L$) of sequential keystroke data. Each sequence is then converted into a two-dimensional grid of specified dimension $L \times B$ (similar to an image), and all sequences are passed as input into the CNN model to be trained, where $L$ is the sequence length (also known as timesteps) and $B$ is the number of features. The overall input size is represented by: samples ($N$), timesteps ($L$), and features ($B$) where samples ($N$) is the total number of sequences created from the dataset. There are two main types of CNN that can be used: the two-dimensional CNN (2D-CNN) and the one-dimensional CNN (1D-CNN). The 2D-CNN is the standard CNN where the kernel (filter) slides along 2 dimensions on the input data and extracts the spatial features \cite{lin2018neural}. The 1D-CNN on the other hand, has a kernel that covers all the features and slides along one dimension \cite{lu2020continuous}. Both the 1D-CNN and 2D-CNN have shown to be effective on time-series sequential data such as the keystroke data, however, the 1D-CNN is preferable and has produced better results in the literature. This is because 1D-CNNs are designed to handle sequences of one-dimensional vectors, while 2D-CNNs are designed to handle images or data represented in two dimensions. In keystroke dynamics, the data is usually represented as sequences of time-based features including but not limited to hold time and inter-key time. 1D-CNNs can capture the temporal patterns in the sequences of these features, whereas 2D-CNNs would not be well-suited for this type of data. Additionally, due to the sequential characteristic of keystroke data extensive padding would be required to produce a 2D representation, which would be computationally expensive and may negatively affect model performance. 1D-CNNs are computationally more efficient and better suited for keystroke dynamics data, as they can effectively handle the sequential patterns in the data without the need for padding. Although CNN is less computationally intensive than a recurrent neural network (RNN), it can not process all keystroke data in a sequence simultaneously and it is unable to capture the sequence order information like RNN would.

\subsubsection{Recurrent Neural Network}
The recurrent neural network (RNN) \cite{mandic2001recurrent} is a deep learning algorithm best used for sequential or time series data such as keystroke data.
An RNN keeps track of information from previous inputs through its memory gate and decides which information in the past should be remembered or forgotten based on the current input and output.
As a result, unlike other deep neural networks, the output of RNN depends on previous information in the sequence, especially for a unidirectional RNN. However, a bidirectional RNN can produce an output that depends on both the past and future information in the sequence. 

\emph{Long Short Term Memory} is a popular RNN architecture, introduced by Hochreiter and Schmidhuber \cite{hochreiter1997long}. It is used to model long-distance relations in input data. The main benefit of LSTM is the inclusion of a cell state, which simply acts like a memory chain by storing information from past states. There are three gates in LSTM: an input gate, an output gate, and a forget gate. The three gates act as logic control within the model used to predict the model's output. The forget gate is computed using equation $f_t = \sigma(W_f \cdot [h_{t-1}, x_t]+b_f)$; where $\sigma$ is the sigmoid function, $h_{t-1}$ is the output from previous step, $x_t$ is the input from current step, and $W_f$ and $b_f$ are the weight matrix and bias for the forget gate. At the input gate, the computations are carried out with these equations $i_t = \sigma(W_i \cdot [h_{t-1}, x_t]+b_i)$; $\hat{C_t} = tanh(W_c \cdot [h_{t-1}, x_t]+b_c)$. Results are then used to update the cell state using equation $C_t = f_t \cdot C_{t-1} + i_t \cdot \hat{C_t}$; where $C_{t-1}$ represents the state of the previous cell. The final output for the current step is calculated using these equations $o_t = \sigma(W_o \cdot [h_{t-1}, x_t]+b_o)$; $h_t = o_t \cdot tanh(C_t)$.

This variant of RNN has been widely used in keystroke dynamics for both authentication and identification \cite{acien2021typenet, kobojek2016application, deb2019actions}. Acien et al. \cite{acien2021typenet} created a deep learning architecture that is quite different from the conventional LSTM architecture with the Siamese Neural Network (SNN) and they called it TypeNet.

SNN is a class of neural network architectures that contain two or more mirrored sub-networks. A major difference between SNN and other traditional LSTM models is that 
an SNN is designed to find the degree of similarity of the inputs to the sub-networks. Hence, it is able to predict new classes of data from users not seen during training without the need to retrain the model. The architecture achieved state-of-the-art performance in keystroke dynamics authentication with EER of 2.2\% and 9.2\% on Aalto desktop \cite{dhakal2018observations} and mobile \cite{palin2019people} datasets, respectively.

\emph{Gated Recurrent Unit} is also a variant of RNN with similar properties to an LSTM \cite{cho2014learning}. Instead of a ``cell state” and three gates used in LSTM, GRU uses hidden states and has two gates: a reset gate and an update gate. The reset and update gates control which information to retain. 
The GRU is without the memory unit which makes it expose the full hidden content without any control. With fewer gates, GRU trains faster, is less complex, and performs better than LSTM, especially on smaller datasets \cite{kobojek2016application, li2022free}. In relation to keystroke dynamics, Kobojek and Saeed \cite{kobojek2016application} contrasted the performance of GRU and LSTM on the same dataset using two scenarios, the first was using only the dwell time, and the second was using all the data. They reported that the GRU (15\% EER) outperformed LSTM (21.9\% EER) when only the dwell time is used, but the LSTM (13.6\% EER) outperformed GRU (22.4\% EER) when all the date was used.

\subsubsection{Transformer}
Stragapede et al. \cite{stragapede2022typeformer} developed a novel Transformer architecture named TypeFormer for mobile keystroke dynamics for user authentication purpose. 
Their Transformer network was based on a two-branch architecture: Temporal and Channel Modules enclosing two Long Short-Term Memory (LSTM) recurrent layers, Gaussian 
Range Encoding (GRE), a multi-head Self-Attention mechanism, and a Block-Recurrent Transformer layer. TypeFormer could map slices of keystroke sequences into a feature embedding space.
Same subject representations of sequences were closer than those of different subjects. The triplet loss function was used to train TypeFormer and the similarity of the feature embeddings
was measured with Euclidean distance. The pre-processed Transformer input sequences were fed to the branches. A GRE was used for modeling the input sequences in both branches; the purpose 
was to preserve the information position. The Temporal Module consisted of three ordered sets of layers. Each of the sets of layers was composed respectively by N, R, and M layers. The N and M
layers were made of two sub-layers: a multi-head Self-Attention mechanism, and a multi-scale keystroke LSTM RNN layer. R recurrent layers were included between the N and M layers. The Channel
Module consisted of H layers similar to the N and M layers of the Temporal Module. There was a convolutional layer at the end of each of the Modules. The output feature embedding vector P was generated by concatenating the similarity of the output features. The output feature embedding vectors were compared through the Euclidean distance for the authentication task. Stragapede used 
TypeFormer on the Aalto mobile keystroke database \cite{palin2019people} and achieved an EER of 3.25\%. 

Neacsu et al. \cite{neacsu2023doublestrokenet} proposed a novel Transformer-based keystroke authentication method named DoubleStrokeNet. In order to generate user embeddings, this model
used the temporal features of bigrams. A Transformer encoder was the main part of the DoubleStrokeNet architecture. The Transformer encoder functioned as a discriminator between legal and
imposter users. It could also differentiate between original and replaced key bigrams. The transformer encoder was able to extract relevant contextual information within data sequences and 
encode key bigrams based on context-aware embeddings. Neacsu et al. used Aalto desktop \cite{dhakal2018observations} and mobile \cite{palin2019people} keystroke datasets for their experiments
and achieved EER values of 0.75\% and 2.35\% for physical and touchscreen keyboards, respectively.

\section{Keystroke Feature Engineering Approaches}
Throughout the following section, the impact of different keystroke feature engineering techniques is described. The performances
of keystroke authentication algorithms depend largely on keystroke data/keystroke features that are used for authentication,
so it is important to determine which features give the best results.

\subsection{Effects of Text Filtering}
Huang et al. \cite{huang2016effects} investigated the effect of gibberish text on the 
performance of keystroke dynamics. Noise or gibberish text can originate when a user 
plays a computer game or does some unnecessary task. Huang et al. decided that the unnecessary text 
should be filtered out due to its negative impact on authentication performance. For the 
evaluation of their hypothesis, they used a novel dataset that was collected when the users 
were performing normal day-to-day activities. There was no restriction or control over
the users. They were free to type anything in order to analyze the noise keystrokes. Huang et al. used a keystroke logger to
collect the data of 60 users in a campus student computer laboratory. Each user provided at least 11000 keystrokes. They used 
Gunnetti and Picardi's algorithm \cite{gunetti2005keystroke} for evaluation. They used samples of 
1000 keystrokes to create a user profile. Each user profile consisted of 10 samples
with the rest reserved for testing. They also used the ``zone-of-acceptance'' algorithm by Leggett et al. \cite{leggett1988verifying}.
From the dataset, they detected four types of gibberish keystrokes: repeating characters, gaming patterns, long strings with few distinct characters, and long strings with no white spaces or separators. They then used two types of 
algorithms in order to remove gibberish keystrokes. The first one was based on regular expressions
that identified four patterns of gibberish keystrokes. The second one was a spell checker
that checked for correct English words. They also used context-based filtering and determined that 23.3\% data in their dataset was gibberish. 
The evaluation showed that though gibberish keystrokes did not have much effect on FAR, 
they negatively impacted FRR. Spell checker filtering showed better performance than regular expressions, but the best result was achieved by combining both filters.

\subsection{Effect of Data Size}
Huang et al. \cite{huang2015effect} analyzed the 
performance degradation of keystroke authentication systems by experimenting with data size. They
analyzed the impact of reference profile size and test sample. They used two algorithms for their 
evaluation process. The first one was Gunetti and Picardi's algorithm \cite{gunetti2005keystroke}. The second one 
was the ``zone-of-acceptance'' algorithm by Legget et al. \cite{leggett1991dynamic}. They used the dataset developed by Vural et al. \cite{vural2014shared}
for their evaluation and performed two experiments. In the first experiment, they detected all instances of 
digraphs and trigraphs in the previously collected user's data, then created a reference profile and a test sample of a specific size.
Every instance of data contained within the reference profile was divided into five samples in order to implement the Gunetti-Picardi algorithm. The dataset contained 
the data of 39 users. They performed the tests 50 times to avoid errors. In the second experiment, each user's data was divided 
into groups of 1000 digraphs. The range of reference profile size was from 6000 digraph instances to 17000 digraph
instances. Only 25 users had enough data to perform the tests. From the first experiment, the results indicated that the FAR value improved when the reference profile size was constant and the test sample size increased. The results also 
showed that keeping the test sample size constant and increasing the reference profile size did not necessarily improve the FAR. 
The IPR (Imposter Pass Rate) improved when the reference profile size increased. The IPR values did not 
decrease for smaller test samples. In the second experiment, Huang et al. used the ``zone-of-acceptance'' algorithm. The results showed that 
both FAR and IPR improved with an increasing the reference profile size. They also found out that the reference profile should consist of a minimum of
10,000 keystrokes or more to achieve good performances. 

\subsection{Impact of Faulty Users' Data}
Ozbek's goal was to improve the classification performance by removing the keystroke data of the users 
who degrade the performance \cite{ozbek2019classification}. Ozbek used two benchmarking databases with different numbers of users and passphrases for the evaluation.
\begin{enumerate}
   \item GREYC Keystroke Data: 133 participants typed the passphrase ``greyc laboratory'' on an AZERTY keyboard for the collection. There were 7555 attempts \cite{giot2009greyc}. For this dataset, Ozbek removed some of the users with a low number of captures leaving only the 100 participants with more than 50 attempts. 
   \item CMU Keystroke Data: 51 participants typed the password ``.tie5Roanl'' with 400 attempts per user \cite{killourhy2009comparing}. 
\end{enumerate}
The data of the two datasets were used as they were and no features were removed. Also, no dimension-reduction techniques were implemented to improve the performance. To evaluate the performance Ozbek used support vector machine (SVM), decision tree, and K-nearest neighbor (KNN) classifiers. For these three classifiers, each keystroke datum was divided into training, testing, and validation sets. 70\% of the data was used for training, 15\%
was used for testing and 15\% was used for validation. No cross-validation technique was applied. Ozbek used a histogram to detect the faulty users. Re-classification of the data was performed with the remaining users. For both datasets,
Ozbek achieved higher accuracy by eliminating the users with misclassifications at the time of training. First Ozbek removed the worst five users which improved the
performance. Ozbek achieved better results by removing more users having misclassifications. However, removing the users from the database was not a standard option.
The main goal was to demonstrate that the worst users could be recognized and eliminated if required.  

\subsection{Impact of Non-Conventional Keystroke Features}
Alsultan, Warwick, and Wei \cite{alsultan2017non} used non-conventional keystroke features for the authentication of users. In the case of the non-convention features,
the focus was mainly on the overall typing patterns. The percentage of performing certain actions was considered, such as editing actions, general typing actions, etc. The goal was
to understand the typing behavior of a user. Two main features were considered - semi-timing and editing features. Semi-timing features were different
from the common timing features used in previous experiments. These features were calculated for greater amounts of time. The semi-timing features that they used were: 1) Word-per-Minute
(WPM) = Number of words / Total typing time in minutes, 2) negUD (negative Up-Down) = Number of negative UDs / Total number of keypairs, 3) negUU (negative UP-Up) = Number of negative UUs / Total number of keypairs. Editing features were characteristics such as typing error frequency, text editing, etc. The editing features
were: 1) Error rate, 2) CapsLock usage, and 3) Shift key usage. Thirty participants provided data for this study and participants had to complete eight typing tasks. The tasks 
included copying the text of 1000 characters. The text was taken from the Guardian newspaper. In the text, there were numbers, both upper and lower case letters, and punctuation marks.
They used a C++ based GUI program for the data collection. The participants could download the application on their own machines. As the user's profile, a feature
vector of nine features was generated and saved in the database. Out of eight typing tasks, each was accomplished with a typing sample from a single source. Features were identified and extracted independently of each typing task. 

In the analysis phase, eight samples per subject were used for training and testing of the classifier. As a classifier, decision trees were used in this
research. Cross-validation was used in the classification process. In their experiment, Alsultan, Warwick, and Wei used eight samples to perform eight cross-validation experiments. They used seven samples for training and one sample for testing. They used the statistics toolbox in Matlab and two error rates to represent the results - False Accept Rate (FAR) and False Reject Rate (FRR). They obtained low error rates in this study and computed the rates of error for a variable number of participants. For the decision tree classifier, they obtained a FAR of 0.007, 0.0104, and 0.0109 for 15, 25, and 30 participants respectively. They obtained a FRR of 0.1, 0.25, and 0.28 for 15, 25, and 30 participants respectively. 

The focus was on true intruder detection in the second part of this study. In this part, typing samples from unknown users were used to test the system. Here, they used binary classification. For the binary classification, they used the training data from 25 legal users and the testing data from five intruders. Each intruder generated three typing samples to test the system. For the decision tree classifier, they achieved a FAR of 0.011 and a FRR of 0.375 for 25 genuine users without any intruders. For the Support Vector Machines (SVMs), they achieved a FAR of 0.0112 and a FRR of 0.49. When the data of the five intruders was tested against the genuine users' data, they got the same FAR in the case of SVMs. In this study, the FRR improved to a value of 0.28 utilizing non-conventional characteristics. The lowest FAR was 0.011, which was also comparable to the results generated by conventional features. 

\section{Keystroke Authentication on Touch Screen and Mobile Devices}
As people are increasingly dependent on touch screen devices such as tablets and mobile phones, it is necessary to implement security and
authentication systems based on these devices. This section describes various aspects of keystroke authentication on touch screen and mobile devices; a summary is shown in Table \ref{table:touchDevice}.

    
\begin{table} [ht]
  \centering
  \scriptsize
  \caption{A summary of keystroke dynamics on touch screen and mobile devices. In this table, algorithms used for keystroke authentication on touch screen and mobile devices are mentioned along with the data on which the algorithms were applied. Also, the results or contributions of the reference papers are provided.}
  \label{table:touchDevice}
  \setlength{\tabcolsep}{5pt}
  \renewcommand{\arraystretch}{2}
  \centering
  \resizebox{\textwidth}{!}{%
  \begin{tabular}{|m{3cm}|m{3cm}|m{4cm}|m{4cm}|}  
    \hline
    		{} & {} & {} & {}\\[-4ex]
        {\textbf{Reference}} & {\textbf{Data}} & {\textbf{Algorithm}} & {\textbf{Result or Contribution}}\\
        \hline
        {El-Abed, Dafer, and Rosenberger \cite{el2018rhu}} & {Password} & {Kruskal-Wallis Test} & {Proof of variation of keystroke feature due to different orientations and different devices}\\
        \hline
        {Gautam and Dawadi \cite{gautam2017keystroke}} & {String} & {Median Vector Proximity} & {Average EER-8.33\% and Standard deviation EER-7.07\%}\\
        \hline
        {Al-Obaidi and Al-Jarrah \cite{al2016statistical}} & {CMU password} & {Statistical Median-Based Classifier} & {EER-6.79\%}\\
        \hline
        {Al-Obaidi and Al-Jarrah \cite{alnoor2016statistical}} & {Password} & {Statistical Distance-to-Median Anomaly Detector} & {EER-0.049\%(for different pass-mark)}\\
        \hline
        {Antal, Szabo, and Laszlo \cite{antal2015keystroke}} & {Password} & {Manhattan Metric} & {EER-15.3\%}\\
        \hline
        {Corpus et al. \cite{corpus2016mobile}} & {Password} & {Neural Network Model} & {Accuracy-73.33\%}\\
        \hline
        {Kambourakis et al. \cite{kambourakis2016introducing}} & {Password and Passphrase} & {Random Forest and k-NN} & {Lowest EER-13.6\%}\\
        \hline
        {Giuffrida et al. \cite{giuffrida2014sensed}} & {Password} & {Machine Learning and Distance Metrics} & {EER-0.08\%}\\
        \hline
        {Lee et al. \cite{lee2018understanding}} & {PIN} & {Distance-Based Classification and OCSVM} & {EER-7.89\%}\\
        \hline
        {Buschek, De Luca, and Alt \cite{buschek2015improving}} & {Password} & {Anomaly Detectors and Classification Methods} & {Combination of spatial and temporal features}\\
        \hline
 \end{tabular}}
\end{table}

\subsection{Impact of Device Models and Orientations}
El-Abed, Dafer, and Rosenberger \cite{el2018rhu} proposed a benchmark for keystroke dynamics using a mobile phone and tablet. The touch screen-based benchmark was
developed in both portrait and landscape orientations. The objective of this benchmark was to evaluate the keystroke authentication algorithms
for varying models of devices and different orientations. An online visualizer was provided to observe the acquired keystroke signals.
They used an Android application for developing their benchmark. They used the Nexus 5 and the Samsung Galaxy Note 10.1 2014 tablet for data collection. There were two main purposes of this keystroke benchmark. Firstly, they wanted to investigate the patterns of typing of the users for a variety of orientations, but the device and the password were the same. Secondly, they wanted to analyze the effect of different devices on the typing patterns of the users, keeping the orientation and the password constant.
Data was gathered from 47 users. The users type the password ``rhu.university'' in four configurations in each session: 
\emph{phone/portrait}, \emph{tablet/portrait}, \emph{phone/landscape}, \emph{tablet/landscape}. The key features of the benchmark
were: \emph{PP}, \emph{PR}, \emph{RP}, \emph{RR}, \emph{TT} (time of typing the password), \emph{Screen Orintation} (portrait or landscape),
and \emph{Screen Size}. They then performed a statistical analysis using their benchmark and used the Kruskal-Wallis test \cite{higgins2004introduction} 
for the analysis. The goal was to analyze the changes in the keystroke features due to different orientations and different devices. The results indicated that
the keystroke features varied significantly for different orientations and different devices. These results must be considered for touch-screen-based mobile authentication technology. 

\subsection{Impact of Different Types of Features}

\subsubsection{Combination of Touch-Based and Timing Features}
Gautam and Dawadi \cite{gautam2017keystroke} analyzed different features of the touch-screen keyboard for keystroke authentication. They combined
the effect of touch-based and timing features and developed a dataset of 7 users. The authentication
system consisted of three steps: Enroll, Verify, and Identify. The user authorization process consisted of two phases. For both the training and enrollment phases, a
reference template was generated. In the testing or authentication phase, the test samples were matched against the reference template using 
the Median Vector Classifier \cite{al2012anomaly}. They used four features for authentication: Key hold time (H), Flight time (FT), Pressure (P), Area (A).
They used the string \emph{.pie7Crawl} for their dataset. The dataset consisted of 47 features. In order to perform their experiment, Gautam and Dawadi utilized a
OnePlus 3 Android phone with an AMOLED capacitive touch screen and 1080 * 1920 pixels (5.5 inches). All 7 participants were touch screen smartphone users.
Every user typed the string \emph{.pie7Crawl7} on the touch screen keyboard. Among the 7 participants, one was considered a legal user, and the other 6 participants were considered imposters. The legal user participated in 40 sessions. Among those sessions, 
10 were reserved for training and the rest for testing. Each of the other 6 participants (imposters) participated in 5 sessions. By using the median vector
proximity algorithm, they achieved an average EER of 8.33\% and a standard deviation EER of 7.07\%. 

\subsubsection{Interaction of Touch Screen Features}
Antal, Szabo, and Laszlo \cite{antal2015keystroke} used their dataset of 42 individuals to investigate the interaction of touchscreen features on the performance of keystroke authentication systems. For data collection, they developed an Android application for a Nexus 7 tablet and a Mobil LG Optimus L7 P710. The application had its own software keyboard. During the registration period, the users had to enter their data, such as their gender, date of birth, experience of smartphone usage, etc. Data was collected in multiple sessions. Most users participated in two sessions 
in two weeks. The users entered the same password (\emph{tie5Roanl}) during each session 30 times. A total of 42 individuals provided their data, with 51 entries for each user.
Entries containing deletions were discarded from the dataset. As all users entered the same password, the data of each user could be used both as a legal user and as an impostor. Among the 42 users, 37 were tablet users and 5 were mobile phone users.
The system saved both the timing and touchscreen features. They used WEKA (version 3.6.11) \cite{ngo2011data} software for their experiment. Weka's search methods were used to optimize some of the default parameters of the classifiers. They used Bayesian network, Nearest neighbors (k-NN, IBk in Weka), Naive Bayes, Decision trees, Support vector machines, Multilayer perceptrons, etc., for the evaluation process. 

In the identification phase, Antal, Szabo, and Laszlo used two datasets to compute the accuracy, one with touch screen features (pressure and finger area) and one without touchscreen features. There were 41 features in the first one and 71 features in the second. The data was used without any modification or feature selection and without boosting or tuning methods. They used 10 fold cross-validation in their experiment and computed the classification accuracies by taking the average of 10 runs. All algorithms performed better on the dataset containing emph{touch screen features}. In the verification phase, an R script developed by Killourhy and Maxion \cite{killourhy2009comparing} was used for performing the measurements. Three anomaly-detection algorithms were included in the R script: Euclidean, Manhattan, and Mahalanobis metrics. The normalized data was split into three equal segments. In each part, there were 17 samples from each user. The user's template was generated using two-thirds of the data, with the remaining data used for FRR testing. In the case of FAR testing, the first five samples from each user were used as impostor data. The measurements were performed three times, using a different fold each time for training and testing. The lowest EER was 12.9\% which was achieved using the Manhattan metric for both timing and touchscreen features. For timing features, the lowest EER was 15.3\%, achieved by the Manhattan metric. Their findings showed that in the identification phase, the accuracy of each classifier increased by 10\% due to the addition of \emph{touch screen features}. In the verification phase, the EER was reduced by 2.4\%.

\subsubsection{Two New Features - Speed and Distance}
Kambourakis et al. \cite{kambourakis2016introducing} developed and evaluated a touchstroke-based authentication system in the Android platform. They used two new features for their experiment apart from common keystroke features viz, speed and distance. They adopted two commonly used authentication procedures. In the first procedure, the users had to type a password, and in the second case,
the users had to type a passphrase. 20 users participated in the data collection between the ages of 19 and 21 years. All users had a touch screen based Android smartphone. 
Kambourakis et al. implemented a prototype of their proposed touchstroke-based authentication system in Google's Android OS. In order to generate the typing behavioral profile of the user and authenticate the user, they used \emph{Waikato Environment for Knowledge Analysis (Weka)} as the classification engine. There were two main analysis sub-systems in their mechanism. 1) Enrollment: Here, the typing profile of a user was created and used to train the classifiers. 2) Authentication: Here, the legitimacy of a user was checked. The typing profile of a user was saved in a database. They considered three machine learning algorithms for their experiment: Random Forest, k-NN, and MLP. However, MLP was rejected due to the memory limitation of smartphones. 

For the password scenario, Kambourakis et al. achieved the best results using the Random Forest classifier and the first data analysis procedure. For the passphrase scenario, k-NN performed better using the second data analysis procedure. For the password scenario, the average FAR\%, FRR\%, and EER\% values were 12.5, 39.4, and 26, respectively. For the passphrase scenario, the average FAR\%, FRR\%, and EER\% were 23.7, 3.5, and 13.6, respectively. They performed their experiment on a Sony Ericsson Xperia ray. The device had a 1 GHz CPU, 512 MB RAM, a 3.3-inch touchscreen, and the Ice Cream Sandwich Android OS. This percentage varied between 91\% and 100\% at the time of training. 
The results showed that the first data analysis procedure needed more memory to perform, as it generated more data than the second one. Both algorithms were able to perform directly on the device, and it took less than one second to authenticate the user.

\subsubsection{Novel Sensor Features}
Giuffrida et al. \cite{giuffrida2014sensed} developed a novel keystroke authentication system enhanced with sensor data for authentication of mobile devices.
They used novel sensor features to analyze the typing behavior of a user. Machine learning techniques were used to authenticate users and to associate the timing-based keystroke features with movement sensor information. They developed UNAGI, a fixed-text authentication system for Android. Their current prototype modified the traditional Android software keyboard and leveraged a number of modules, e.g., feature extraction, training, and detection. The sensor-enhanced keystroke dynamics
techniques were implemented through these support modules. The user had to type a fixed-text password during an authentication session. Their authentication system processed this
password for analysis. When the user typed, UNAGI recorded the key-press events and also sampled movement sensor data periodically. They used two Android sensor sampling interfaces:
TYPE\_LINEAR\_ACCELERATION and TYPE\_GYROSCOPE. Sensor values were collected at a high sampling frequency. In order to accomplish this, the SENSOR\_DELAY\_FASTEST flag was specified at
sensor listener registration time. The feature extraction module of UNAGI processed the collected data and extracted suitable features.
Features from the training sessions were fed to the training module to develop a user profile. The detection module compared the features of a testing session against the user profiles for authentication of genuine users. 
The movement sensor features considered by UNAGI were root mean square, number of local maxima and minima, minimal and maximal value, mean delta, the sum of negative values, the sum of positive values, mean value, mean value during KD and KU events, and standard deviation. The keystroke features considered by UNAGI were KD-KD time, KD-KU time, KU-KD time, and KU-KU time. After the feature extraction phase, the output was a vector that consisted of all the features, both keystroke and movement sensor features. In their experiment, threshold-based binary classification algorithms were suitable.
Giuffrida et al. used one-class SVM, Naive Bayes, k-Nearest Neighbours (kNN), and the mean algorithm. They also used several distance metrics - Euclidean, Euclidean normed, Manhattan, Manhattan scaled, and Mahalanobis. They used three different configurations for evaluation purposes: keystroke data only, sensor data only, and a combination of both. Results showed better accuracy 
with sensor-based features rather than keystroke timing features, i.e., an EER of 0.08\% for sensor-based features vs. an EER of 4.97\% for keystroke timing features.

\subsubsection{Touch-Based Features with Three Distinct Hand Postures}
Buschek, De Luca, and Alt \cite{buschek2015improving} conducted an experiment investigating mobile keystroke biometrics. They compared touch-based features in consideration of three distinct hand postures and evaluated touch-specific features for user authentication. They used two types of typing features: 
\begin{enumerate}
 \item Temporal features: hold time, flight time, up-up times, down-down times, etc. 
 \item Spatial touch-specific features: exact touch locations, offsets, touch jumps, drag, touch area sizes, ellipse axes, touch pressure, etc. 
\end{enumerate}
They used two types of models: 1) anomaly detectors, and 2) classification methods. For this study, data were collected in two sessions with a one-week gap between sessions. A total
28 users participated in the data collection, and the average age of the users was 25. 

A Nexus 5 phone was utilized for the data collection. A unique
software-based keyboard was used to extract the features. Each user participated in two sessions. Each session included data collected using three separate hand orientations. The participants were instructed to type 6 distinct passwords 20 times using each hand posture. From their analysis, it was observed that models achieved better results on data from a
single session - the EERs were less than half of the values obtained for data across sessions. In the case of evaluation, classifiers performed better than anomaly detectors
(11.7-31.1\% lower EERs). EERs increased by 86.3\% for different hand postures. Regarding the features, the best spatial feature sets performed better than the best temporal sets.
EERs were reduced by 14.3 - 23.5\% using the spatial touch features relative to the traditional temporal features. Also, EERs were reduced by using spatial touch features instead of
pressure features. The combination of spatial and temporal features provided the best results; EERs were reduced by 8.5 - 26.3\% relative to the best spatial features and by 26.4 - 36.8\% relative to the best temporal features.

\subsection{Fusion of Keystroke Dynamics and Other Mobile Modalities}
Corpus et al. \cite{corpus2016mobile} described the use of accelerometer biometrics (the measurement of how 
a person holds their mobile device) in conjunction with keystroke dynamics. They proposed a model for security in mobile applications by combining accelerometer biometrics with previously studied keystroke dynamics.
The goal was to apply their model in a real-world use case. For data collection, they created an application using Android Studio 1.3.2. They used a QWERTY soft keyboard and 
collected the keystroke data from 30 users. Users had to type a password of 8-16 characters a total of 8 times. The collected data were stored in two different text files. The first 
file recorded the data related to keystroke dynamics, and the second file contained the data on accelerometer biometrics. 

Among the 8 samples of each user, the first 6 samples 
were used for training and the remaining 2 samples were used for testing. Their models were: Naive Bayes, Decision Tree, J48, and Neural Networks. They developed their models on three sets of 
features to train their models - accelerometer biometrics only, keystroke features only, and lastly, a combination of both. The neural network model showed the best performance for 
combined features. However, adding feature selection improved the performance. Chi Square attribute was used to rank the features, and the low-rank features were removed.
Corpus et al. also removed features that were highly correlated with other features. After data cleaning, the accuracy increased to 68.89\% by implementing the feature selection techniques. They used a pre-labeled test set to test the best model. The test accuracy was 73.33\%, FAR was 27.59\% and FRR was 26.67\%. Corpus et al. modified their data
collection tool for real-world testing through a prototype. The enrollment of the username and four password samples was included in the mobile application. A web server was used
to implement re-modeling and testing processes. Six users participated in this test. There was no password sample for these six users
in the database. Each user chose their own password and typed the password four times. The users tested their own passwords and the passwords of their co-participants. For the prototype 
combining accelerometer biometrics with keystroke dynamics, they concluded that accelerometer biometrics has an impact on keystroke dynamics-based authentication. Though the 
prototype showed low FAR (7\%), the recognition performance was above average (60\%-70\%). It concluded that the prototype was not fully accurate and ready to deploy.  

Lee et al. \cite{lee2018understanding} analyzed smartphone-based keystroke dynamics using a 6-digit pin. They used a distance-based classification algorithm
and extracted two types of smartphone data, motion and keystroke data. The extracted keystroke features included the time of events, size of the fingertip,
coordinates of the touch point, etc. The extracted motion features included accelerometer, rotation, gravity, etc. The raw data was collected by typing 
the 6-digit PIN - ``766420''. The app was installed on a Nexus 5X. Data was collected from 22 users with 100 samples from each user. One sample
contained a one-time input of a 6-digit PIN. They used two algorithms for their experiments, distance-based classification and OCSVM (one-class SVM). In the case of distance-based classification,
two scaling methods were used: MinMax scaling and standard scaling. Two distance metrics were used: Euclidean distance and Manhattan distance. The distance-based classification experiments had two results:
\begin{enumerate}
 \item Adding features from motion data: Lee et al. used five different formulas featuring motion data for their experiment. The best result was achieved using standard scaling and Manhattan distance. The lowest EER was 12.63\%. Lee et al. obtained an EER of 8.94\% using only
the keystroke data. The EER was decreased by 1.05\% to 7.89\% when the motion information was introduced. 
 \item FAR by Gender Match: It was observed from the result that the opposite gender's match had an impact on FAR. A lower FAR was obtained if the gender of a genuine user and the gender of an impostor were different. For male genuine users, the FAR decreased by 4.07\% in the case of different genders. For female genuine users, the FAR decreased by 0.64\% in the case of male impostors. 
\end{enumerate}
In the case of one-class support vector machines (OCSVMs), they computed the results for keystroke samples only and after adding motion data. For OCSVM, the EER decreased by 1.24\% adding the motion data.
In the case of the experiment about gender match using the OCSVM, FAR decreased by 8\% for male genuine users and FAR decreased by 6\% for the other cases.

\section{Other Applications of Keystroke Dynamics}
The number of applications that rely on input from a computer keyboard
is increasing as computers have become more commonplace in
the home, workplace, and academia. People have become more comfortable with
typing on keyboards. In fact, today’s self-taught typists type almost as fast as
touch typists \cite{logan2016different}. This increase in the number of people who are
typing on a keyboard creates additional opportunities to apply
keystroke dynamics to domains not previously envisioned. An authentication API for adopting keystroke authentication is
\href{https://www.typingdna.com/authentication-api.html#demo}{typingdna}.
This keystroke authentication API can be used to secure logins, enforce password resets, and authenticate subjects. 

The examples mentioned in this section are examples of
domains where fixed-text and free-text keystroke dynamic analysis
can be applied.

\subsection{Educational Level Classification}
Tsimperidis et al. \cite{tsimperidis2018r} used keystroke dynamics for the detection of one's level of education. They proposed the randomized radial basis function 
network, which is a novel machine learning model. Their model is capable of identifying the educational level of the keyboard user. Their model is 
the first model leveraging keystroke biometrics for the detection of one's education level. Their experiment was performed in two steps. First, free-text unrestricted data were collected from the participants for ten months. The second step was developing the R\textsuperscript{2}BN model. The computational 
time of the R\textsuperscript{2}BN model was very high. In order to reduce that time, they decreased the size of the dataset without degrading the 
performance of the model. A filter-based random subfield data condensation approach was proposed. 242 log files were collected from five educational-level classes. There were 2800 to 4500 keystrokes in each log file. 162 features were extracted from the data. The data was fed
into the R\textsuperscript{2}BN model. This model was developed by the randomized modification of a popular radial basis function neural network (RBFN), proposed by Broomhead and Lowe \cite{broomhead1988multivariable}. The advantage of R\textsuperscript{2}BN model was its speedy convergence and smaller extrapolation errors due to a shallow-learning architecture. This model achieved more than 85\% accuracy in the educational level classification task using keystroke dynamics. 

\subsection{Emotion Detection}
Detecting emotions from your counterpart during electronic
communications, such as email or chat, is often difficult. Written
words can be misinterpreted into thinking the other side is angry, or
conversely, saddened, when in reality, the real emotion behind the
discussion is unknown. Determining the emotion associated with the
text can lead to improved communication. Feelings such
as anger, sadness, disgust, fatigue, joy, fear, and indifference can be
assumed from the way a user types \cite{nahin2014identifying}. Associating emotions based
on the way an individual types can reduce guesswork on the
emotional state of the author and reduce confusion. Such a system
can also serve as a warning for users to rethink or delay impulsively
sending emails or other messages when angry, preventing regret.

Qi, Jia, and Gao \cite{qi2021emotion} proposed an emotion recognition system based on piezoelectric keystroke dynamics.
Piezoelectric materials are able to reflect the force detection sensitivity. Their experiment was simply a password entry. They used time and pressure dimension features.
For emotion recognition, they utilized the discrete PAD 3-dimensional model. International Affective Digitized Sounds (IADS) (a standardized set of 110 digitalized sounds) was implemented to reveal emotions. In order to measure the amount of emotion induction, they implemented a PAD emotion scale. Their target was to classify 4 emotions - happiness, fear, sadness, and disgust. They obtained an accuracy of 78.31\% using the Random Forest Classifier. Their proposed approach proved to be a pragmatic one for emotion classification in practical applications. 

Nahin et al. \cite{nahin2014identifying} identified the emotions of users by analyzing the textual patterns of the user's keystroke dynamics. They considered seven basic emotional classes, viz. anger, guilt, disgust, joy, fear, shame, and sadness. They used the categories from the International Survey on Emotion Antecedents and Reactions (ISEAR) \cite{scherer1990international}. There were two steps in their data collection process. 1) A Java-based software to collect fixed-text data, 2) An app written in C\# to collect free-text data. After data collection, a software program written in C was used for keystroke feature extraction from the log files in order to separate feature files. 19 keystroke 
features were extracted from the collected data. Nahin et al. used the best seven features for analysis. Weka 3.6.6 was used for training. They developed different models for fixed-text and free-text data. For text pattern analysis, the ISEAR dataset was used \cite{scherer1994evidence}. In order to determine emotion class from raw text, they used VSM or term vector model \cite{salton1975vector}. For fixed-text analysis, the Weka tool was used \cite{witten2005practical}. These models were used to evaluate test datasets. They used these simple logistics, SMO, Multilayer Perceptron, Random Tree, J48, and BF Tree for evaluation. They used these same methods for free-text analysis. They calculated the accuracies of emotion class recognition for both fixed-text and free-text. In the case of VSM, the raw data were used as input. VSM identified the emotional state of a sentence by analyzing the text. In the case of the final result, only those results were considered right that were the same for both keystroke dynamics and text pattern analysis. The combined results showed better accuracies than the separate results. 

\subsection{Age, Gender, and Demographics Prediction}

\subsubsection{Predicting Age and Gender}
Pentel \cite{pentel2017predicting} used keystroke dynamics and mouse patterns to predict the characteristics of the user, including gender and age. Pentel collected data from six sources for this study between 2011-2017. To collect mouse movement events, a website that recorded mouse events was leveraged. To collect keystroke data, a key logging system was used, which was written in JavaScript. 
Pentel used five machine-learning algorithms implemented in Java, viz. Logistic Regression, Nearest Neighbor,
Support Vector Machine, Random Forest, and C4.5. In the experiment, Pentel performed the classification of 10-15 years old separate from the other users in order to 1) balance groups with less under-sampling, 2) compare n-graph features based on time with features based on
frequency, and 3) juridical purpose. For evaluation 10 fold cross-validation was used on all models. Pentel performed the training and validation 10 times. A single f-score value was used to present the results, which were calculated by taking the weighted average of both classes' f-score values. Two sample T-tests were performed between each age group to determine the variation between groups in typing speeds. Pentel computed the results of gender and
age-based classification. For each classification, the baseline was 0.5. All the results were over baseline. For keyboard features, the best f-score for gender-based classification was
0.73, and the best f-score for age-based classification was also 0.73. Pentel achieved the best results using Random Forest for keyboard features. However, models trained on
mouse features showed better results due to more mouse instances. Significant differences in typing speed were found between different age groups. T-test results showed that typing 
of the age group 16-29 was faster than other groups. Pentel compared the best time-based features with the frequency-based features of previous studies, and there was some overlap
between the two sets. 

\subsubsection{Gender Recognition}
Tsimperidis, Arampatzis, and Karakos \cite{tsimperidis2018keystroke} used keystroke dynamics to determine the gender of a user. They created a new dataset to identify the most useful features for gender recognition. There were three phases in their experiments. During the first phase, participants provided free-text data. 
In the second phase, features were selected via a selection algorithm. The features were sorted based on the information they contained. In the third phase, the results obtained
by using different algorithms were compared. A free-text keylogger, called IRecU, was created for data collection. This keylogger was suitable for any Microsoft
Windows-based devices. The participants had to install IRecU on their devices and use the keylogger during their
daily activities without any restriction. 75 volunteers provided the log files containing the data. Data was collected over a period of 10 months. The dataset
consisted of 248 log files; 125 labeled as ``male'' and 123 labeled as ``female''. There was data from 2800 to 4500 keystrokes
in the log files. 

A software application named ISqueezeU was developed for feature extraction from the log files. The software application read the text files of the keylogger and then
computed the average number of keystroke timings, including keystroke durations and down-down digraph latencies. They only considered the keys that appeared at least 5 times and the digraphs that appeared at least 3 times. Over 10000 features were extracted. 
Due to the number of features, an automated selection method was required. The models that they used were: SVM, RF, NB, RBFN, and MLP. In their experiments, they tried to determine the gender of a user only by typing speed. The mean of all digraph latencies in each log file was computed, as it represented the typing speed of a user. Then they calculated and compared the 
male and female average values. For males, the average value of all digraph latencies was 373.04 ms, and the standard deviation was 135.26 ms. For females, the average value was 375.71 ms, and the standard deviation 
was 116.86 ms. As the values were very close for males and females, recognizing the gender of a user using typing speed was not possible. Keystroke dynamics dataset and 
several different sets of features were used in order to evaluate the performance of the machine-learning models, namely random forest (RF), support vector machine (SVM), 
naive Bayes (NB) classifier, multi-layer perceptron (MLP), and radial basis function network (RBFN). The calculated metrics were: the model accuracy (Acc.), the
time complexity (TBM), the F1-score (F1), and the ROC index (AUC). The goal was to find out the optimal set of features. They performed numerous experiments on the five 
models mentioned earlier. Each time, they used a variable number of keystroke features. In every case, they used the polynomial kernel as the kernel type because of its better
performance. Neural networks and SVMs performed better for gender recognition using features of keystroke dynamics. However, RBFN performed even better. They drew three conclusions based on their experimental results. 
\begin{enumerate}
 \item For all models, the highest accuracy was achieved before using the highest number of keystroke dynamics features. 
 \item In the range of 150-350 features, the five tested models achieved almost constant accuracy. 
 \item The RBFN model achieved 95.6\% prediction accuracy for 350 features, which was the highest gender prediction rate using keystroke dynamics. 
\end{enumerate}

\subsubsection{Predicting Typist Cognition and Demographics}
Brizan et al. \cite{brizan2015utilizing} used features based on keystroke dynamics, stylometry, and ``language production'' in order to recognize the cognition and demographics
of a typist. For their task, they used a dataset consisting of 350 users. The dataset contained free-text data. 1013 participants were included in the data collection phase for a total of two phases. 838 users participated in the first session, 491 users participated in the second session, and 486 users attended both sessions. Each user was provided a unique ID to identify them. 

Demographics of age range, native language, gender, typing style, and handedness were collected.
The features that they used were split into 3 categories: 
stylometry, language production, and keystroke dynamics, leading to a total of 2381 features. The keystroke dynamics features that they analyzed were user typing speed,
durations and frequency of pauses, pauses before specific keys, the average latency between any two keystrokes, key hold, etc. The analyzed stylometric features were linguistic
unit lengths, character type, consonant frequency, lexical diversity, lexical density, etc. The language production features were part-of-speech pauses, punctuation pauses,
misspelling pauses, revision features, typing burst features, etc. 
They performed two sets of experiments.
\begin{itemize}
\item Cognitive Task Prediction: Four experiments were performed to identify the cognitive task. Four classifiers were used to predict the task type: SVM with a Linear Kernel, SVM with an RBF Kernel, Naive Bayes, and AdaBoost with single split decision trees. The results showed that the SVM classifier performed better 
than other classifiers in predicting the cognitive task. There was no significant difference due to the selection of kernel - RBF or Linear. From the results, it could be 
suggested that it was possible to predict the type of task that a user was performing based on the short analysis of the typing behavior of that user. 
The results of the binary classification showed some minor differences between the low cognitive demand versus the high cognitive demand.
\item Prediction of demography: Several experiments were performed to predict handedness, gender, and native language. They extracted the demographic labels based on self-reports of the users. The experiments were performed on training and test data. They used four classifiers for their experiments: LogicBoost, Naive Bayes, SMO (RBFKernel), and SimpleLogistic. The Simple Logistic classifier performed better than other classifiers for gender classification. The Naive Bayes classifier generated the most promising results for native language and handedness prediction.
The demographic prediction results did not vary due to cognitive tasks. Prediction results were computed for each answer. 
They were able to accurately predict all 3 demographics for 55\% of the answers. A Minimum of 2 of the 3 demographics were correctly predicted for 95\% of the answers. Among
the features, three features were very helpful for gender classification: 1) the timing between punctuation and hitting the spacebar, 2) the timing before and after
function keys, and 3) the timing before and after common digraphs. It was observed that non-native English speakers typed more slowly than native English speakers. Mainly, 
three features were important for primary language prediction: 1) the timing before and after function keys, 2) the timing before and after the `.' key, and 3) the timing
before and after common digraphs. 
\end{itemize}

\subsection{Disease Detection}
It is desirable to use keystroke dynamics for self-reporting, and self-monitoring for medical conditions, to include capturing indicators
of declining health. The purpose is to identify and detect changes in
typing behavior believed to be the same user, but changes in typing
patterns suggest testing for health risks. Parkinson’s Disease is a
highly misdiagnosed, progressive neurodegenerative movement
disease that exhibits a change in motor function over time, which
can be measured in the fingers. Research has been conducted for
the early detection of Parkinson’s Disease by detecting changes in the
characteristics of finger movements typed on a keyboard to classify
people with Parkinson’s from those without the disease. Patients
without Parkinson’s disease were distinguished from those who were
diagnosed \cite{adams2017high}. It was also theorized that a simpler model to classify keystroke variances can be effective while reducing computing resources and requiring fewer keystrokes for training, making such detection systems practicable \cite{milne2018less}.

\subsection{Mental Fatigue Detection}
Conducting intricate tasks requires attentiveness.
Fatigue contributes to errors and mistakes. Long-haul truck drivers who have worked long hours often suffer from an increase in fatigue-related
accidents. Drivers are now required to comply with Electronic
Logging Devices, or ELD, to limit the time on the road in an attempt
to reduce accidents. Electronic logbooks are congressionally
mandated and intended to help create a safer work environment for
drivers. An ELD synchronizes with a vehicle engine to automatically
record driving time for easier and more accurate hours of service
recording. Similar to driving risks, other professions, such as
surgeons, first responders, and pilots, can also benefit from fatigue
detection. Instead of limiting productivity by
time, dangerous levels of fatigue can be detected by the way a user
types using Keystroke Dynamics \cite{acien2022detection}.

\section{Research Opportunities}
The current landscape of keystroke dynamics authentication systems presents several limitations and challenges that need to be addressed to enhance security and user experience.
Firstly, some conservative users may find keystroke dynamics authentication inconvenient, which may lead to reluctance in its adoption. Additionally, these systems are vulnerable to spoofing attacks in which synthetic keystroke dynamics samples can be forged and maliciously used to compromise security. Furthermore, various factors such as noise, different keyboards, typing on unfamiliar devices, physical condition, or urgency can affect a user's typing pattern or speed. Therefore, these factors need to be considered seriously when designing keystroke dynamics-based systems. Moreover, ensuring the privacy and security of users' keystroke data is paramount.

To overcome these challenges and improve keystroke dynamics-based systems, several research directions can be explored.

\begin{itemize}

\item\textbf{Research Benchmarks}: Shared research benchmarks are necessary to truly understand the strengths and limitations of each algorithm and compare them fairly and reliably.  For example,  Stragapede et al. present a new experimental framework to benchmark keystroke dynamics-based biometric verification performance and fairness~\cite{stragapede2023keystroke}.
\item \textbf{Adaptive Authentication Systems}: Researchers can focus on developing novel keystroke authentication systems that can \emph{adapt} to the changing typing behaviors of users in various circumstances. This adaptability can significantly improve authentication performance by accommodating variations in typing patterns.
\item \textbf{Optimization of Training Procedures}: In order to optimize the training procedure of keystroke authentication models, an embedding vector model can be learned as commonly done in the computer vision field. A much larger dataset of keystroke sequences would need to be collected. The model will be trained using keystroke sequences as input. The learned model will convert each keystroke sequence or feature set into an embedding vector representation. For this procedure, neural network architectures can be adopted, such as recurrent neural network (RNN), convolutional neural network (CNN), and transformer.  
\item \textbf{Addressing Synthetic Forgeries}: Synthetic forgeries of keystroke dynamics samples are a potential research area.  Gonzalez \cite{gonzalez2023ksdsld} introduced various innovative techniques to generate counterfeit keystroke dynamics samples using genuine free-text samples provided by legitimate users. In addition, a liveness detection approach was proposed that could differentiate between human-written samples and synthetic forgeries. Gonzalez et al. \cite{gonzalez2022towards} presented two approaches to create synthetic keystroke timing samples, as well as a liveness detection system for keystroke dynamics that employed these synthetic samples as adversaries.
\item \textbf{Employment of Cryptographic Techniques}: To ensure the privacy and security of users' keystroke data, cryptographic techniques can be employed. These techniques can include encryption, hashing, and secure transmission protocols to safeguard sensitive keystroke data from unauthorized access or tampering.
\item \textbf{Integration with Other Biometric Modalities}: To increase the robustness of authentication, keystroke dynamics can be combined with other biometric modalities such as face or fingerprint recognition. This multi-modal approach would enhance security and accuracy by adding additional layers of authentication.
\end{itemize}

By addressing these research opportunities, keystroke dynamics-based authentication systems can become more robust, secure, and user-friendly, thus fostering greater trust and adoption among users.

\section{Conclusion}
Keystroke dynamics is a very effective form of behavioral biometrics and remains an area of active research. 
This survey reviews various aspects of keystroke dynamics, which can be very helpful for future research in this domain.
Keystroke dynamics can be a very powerful tool for security and authentication, a significant reason for which is that the application and deployment of keystroke dynamics-based systems is cost-effective and user-friendly.
This survey covers the state of the art of keystroke dynamics research and applications.
This includes the most important keystroke datasets as well as keystroke authentication algorithms.
Different keystroke feature engineering approaches like text filtering, data size variation, removing faulty data, and using non-conventional keystroke features are described.
It also covers aspects of keystroke authentication on touch screen and mobile devices.
Finally, various applications of keystroke dynamics are described in detail.

\section{Acknowledgments}
This material is based upon work supported by the Center for Identification Technology Research and the National Science Foundation under Grant No. 2413228 as well as 2122746 and 2526924.

\bibliographystyle{plain} 
\small{
 \bibliography{refks}
}
\end{document}